%% file: main.tex
\documentclass[letterpaper,twocolumn,10pt]{article}
\usepackage{usenix-2020-09}
\usepackage{pifont}
\usepackage{balance}
\usepackage{enumitem}
\usepackage{graphicx}
\usepackage{subfigure}
\usepackage{algorithm}
\usepackage{tablefootnote}
\usepackage{threeparttable}
\usepackage{xspace}

\usepackage{algorithmicx}
\usepackage{textcomp}
 \usepackage{algpseudocode}
 \algdef{SE}[DOWHILE]{Do}{doWhile}{\algorithmicdo}[1]{\algorithmicwhile\ #1}%

\algrenewcommand\algorithmicrequire{\textbf{Input:}}
\algrenewcommand\algorithmicensure{\textbf{Output:}}

\usepackage{amsmath}
\usepackage{amsfonts}
\usepackage{amssymb}
\usepackage{bm}

\usepackage{graphicx}
\usepackage{tabularx}
\usepackage{threeparttable}
\usepackage{makecell}
\usepackage{multicol}
\usepackage{multirow}
\usepackage{xparse}
\usepackage{arydshln}

\def\ie{\textit{i.e.}\xspace}

\def\etal{\textit{et al.}\xspace}

\def\eg{\textit{e.g.}\xspace}

\newcommand{\rednote}[1]{\textcolor{black}{#1}}
\newcommand{\bluenote}[1]{\textcolor{black}{#1}}

\usepackage{tikz}
\usepackage{amsmath}

\usepackage{filecontents}

\begin{document}

\date{}

\title{\Large \bf Adaptive Parameter-Efficient Federated Fine-Tuning on Heterogeneous Devices}
\author{
Jun Liu$^{1,2}$, Yunming Liao$^{1,2}$, Hongli Xu$^{1,2}$, Yang Xu$^{1,2}$,
Jianchun Liu$^{1,2}$, Chen Qian$^{3}$\\
$^{1}$School of Computer Science and Technology, University of Science and Technology of China\\
$^{2}$Suzhou Institute for Advanced Research, University of Science and Technology of China\\
$^{3}$Department of Computer Science and Engineering, University of California at Santa Cruz\\
}




    

\maketitle

\begin{abstract}
\input{contents/abstract}
\end{abstract}

\section{Introduction}
\input{contents/intro}


\section{Background and Motivation}\label{sec:prelim}
\input{contents/prelim.tex}

\section{System Overview}\label{sec:algorithm}
\input{contents/overview}

\section{{System Design}}\label{sec:design}
\input{contents/design.tex}


\section{Implementation}\label{sec:implementation}
\input{contents/implementation}

\section{Evaluation}\label{sec:evaluation}
\input{contents/simulation.tex}
\section{Related Works}\label{sec:relatedwork}
\input{contents/related_word}


\section{Conclusion}\label{sec:concluesion}
\input{contents/conclusion}

\balance
\bibliographystyle{unsrt}
\bibliography{contents/refs}

\end{document}

%% file: contents/abstract.tex
Federated fine-tuning (FedFT) has been proposed to fine-tune the pre-trained language models in a distributed manner.
However, there are two critical challenges for efficient FedFT in practical applications, \ie, resource constraints and system heterogeneity.
Existing works rely on parameter-efficient fine-tuning methods, \eg, low-rank adaptation (LoRA \footnote{Note that LoRA in this context refers to not the long-range radio communication technique, but the deep learning fine-tuning technique.}), but with major limitations.
Herein, based on the inherent characteristics of FedFT, we observe that LoRA layers with higher ranks added close to the output help to save resource consumption while achieving comparable fine-tuning performance.
Then we propose a novel LoRA-based FedFT framework, termed LEGEND, which faces the difficulty of determining the number of LoRA layers (called, LoRA depth) and the rank
of each LoRA layer (called, rank distribution). 
We analyze the coupled relationship between LoRA depth and rank distribution, and design an efficient LoRA configuration algorithm for heterogeneous devices, thereby promoting fine-tuning efficiency. 
Extensive experiments are conducted on a physical platform with 80 commercial devices. The results show that LEGEND can achieve a speedup of 1.5-2.8$\times$ and save communication costs by about 42.3\% when achieving the target accuracy, compared to the advanced solutions.

%% file: contents/intro.tex
\vspace{-0.2cm}
The emergence of transformer \cite{vaswani2017attention} and its variants \cite{bai2020binarybert, devlin2018bert, hou2020dynabert} has catalyzed significant advancements in natural language processing (NLP), unveiling the immense potential for deploying NLP models on various devices.
The existing NLP paradigm encompasses two stages, \ie, pre-training and fine-tuning \cite{brown2020language, devlin2018bert}. 
Specifically, the language model (LM) is first pre-trained on a large corpus to learn general features and patterns. 
Subsequently, the LM is further fine-tuned on domain-specific data generated on devices to enhance the model performance for a specific task. 
However, it is infeasible to collect enough domain-specific data from devices for centralized fine-tuning due to data privacy \cite{mcmahan2017communication, lin2021fednlp, liao2023accelerating}. 
To fully utilize the massive data on devices, federated fine-tuning (FedFT) has been proposed to perform fine-tuning in a distributed manner \cite{lin2021fednlp}.
In the typical FedFT framework, \eg, FedNLP \cite{lin2021fednlp}, participating devices periodically fine-tune the LMs on their local data, and push the local LMs to the parameter server (PS) for global aggregation without exposing their raw data.
The fine-tuning procedure is repeated for multiple rounds until the LM converges or reaches the target accuracy \cite{mcmahan2017communication, stremmel2021pretraining, cai2022fedadapter}.
FedFT not only protects individual privacy but also fully utilizes plenty of computing resources on devices to enhance the fine-tuning performance of the LMs.

\textbf{Challenges of FedFT.}
Although FedFT has demonstrated its advantages, it still faces the following two challenges in practical applications: 
(1) \textit{Resource constraints}. 
Many devices such as smartphones and in-vehicle devices typically have limited resources (\eg, memory, computing power), which are orders of magnitude weaker than cloud servers \cite{xu2022adaptive, zhu2023pockengine, dhar2021survey, liao2024mergesfl}. 
However, existing LMs, \eg, Llama \cite{touvron2023llama}, typically involve billions of parameters, requiring substantial computing power for fine-tuning \cite{hu2021lora}, while resource-constrained devices always lead to very slow fine-tuning rates.
(2) \textit{System heterogeneity}.
The devices commonly possess varying computing capabilities (\eg, CPU frequency) or communication capabilities (\eg, bandwidth), which could differ from each other by more than tenfold \cite{liao2023adaptive, lai2021oort, liu2023yoga}.
Specifically, there are huge gaps of computing/communication capabilities among different types of devices, and even among devices of the same type with diverse configurations (\eg, smartphones, laptops) \cite{dhar2021survey}.
Due to system heterogeneity, fast devices should be forced to wait for slow ones, leading to prolonged waiting time and poor fine-tuning efficiency.

\textbf{Status Quo and Limitations.} 
To handle the issue of resource constraints, existing works rely on parameter-efficient fine-tuning methods \cite{zhang2023fedpetuning}, \eg, Adapter \cite{houlsby2019parameter} and low-rank adaptation (LoRA) \cite{hu2021lora}, which only fine-tune additional lightweight parameters (typically less than 1\%). 
The Adapter method inserts additional blocks between two continuous transformer layers and only updates the parameters of the inserted blocks to achieve efficient fine-tuning \cite{houlsby2019parameter}.
For example, Cai \etal \cite{cai2022fedadapter} first apply Adapter in FedFT and propose FedAdapter, which dynamically searches for the optimal Adapter structure to improve the fine-tuning efficiency.
However, the Adapter method inevitably brings additional inference latency, potentially resulting in up to a 30\% latency increase \cite{hu2021lora, liao2023parameter, li2024caraserve}, which is often unacceptable in practical applications, \eg, real-time sentiment analysis \cite{wankhade2022survey} and real-time news categorization \cite{minaee2021deep}. 
In addition, none of these methods have been tested under real wireless network environments with heterogeneous devices.

To avoid the extra inference latency, LoRA \cite{hu2021lora} has been proposed to freeze the pre-trained LM and add trainable bypass low-rank matrices (\ie, \textit{LoRA layers}) to the transformer layers, in which only the LoRA layers would be updated during fine-tuning.
The vanilla LoRA adds the LoRA layers with the \textit{same} rank (\ie, the dimension of the bypass low-rank matrices) to \textit{all} transformer layers \cite{zhang2022adaptive}.
To explore the potential of LoRA in FedFT, Zhang \etal \cite{zhang2023fedpetuning} propose FedLoRA and verify the efficiency of FedFT with vanilla LoRA through experiments.
Upon FedLoRA, Cho \etal \cite{cho2023heterogeneous} propose HetLoRA, in which each device adds LoRA layers to all transformer layers with a diverse and appropriate LoRA rank to deal with system heterogeneity.
However, due to the rank mismatch of all LoRA layers on different devices, it is difficult to aggregate these layers, resulting in poor fine-tuning performance.
In a nutshell, existing works simply add LoRA layers with the uniform rank distribution for all transformer layers, which still requires substantial computing and communication resources, resulting in slow fine-tuning speeds on weak devices.
Moreover, system heterogeneity further leads to low fine-tuning efficiency or poor fine-tuning performance \cite{kim2021autofl, luo2022tackling, liao2024mergesfl}. 
They do not address the two above challenges. 

\textbf{Overview of the Proposed Approach.} 
According to the pre-test in Section \ref{sec:prelim}, adding the LoRA layers with different LoRA configurations (\eg, LoRA depth and rank distribution) significantly impacts the fine-tuning performance and resource consumption, where LoRA depth denotes the number of the continuous LoRA layers from the output. 
Adding LoRA layers with higher ranks to transformer layers close to the output helps to save resource consumption while achieving comparable fine-tuning performance.
Based on the insight, we propose a novel FedFT framework with adaptive \textbf{L}oRA depth and rank distribution on h\textbf{e}tero\textbf{gen}eous \textbf{d}evices, termed LEGEND, to deal with resource constraint and system heterogeneity.
Our unique findings include: 1) 
large LoRA depth with high rank helps to achieve superior fine-tuning performance but incurs significant resource consumption, leading to a slow convergence rate on devices with limited resources; 2) Small LoRA depth with low ranks reduces the fine-tuning overhead on devices while resulting in poor fine-tuning performance or even failure to converge.
Therefore, it is necessary yet challenging to simultaneously determine the appropriate LoRA depth with reasonable rank distribution for heterogeneous devices, so as to well balance the trade-off between fine-tuning performance and resource consumption.
To our knowledge, this is the first study of federated learning based NLP in a real wireless testbed with heterogeneous devices.
The main contributions of this paper can be summarized as:
\vspace{-0.1cm}
\begin{itemize}
    \vspace{-0.2cm}
    \item To address the challenges of resource constraints and system heterogeneity, we propose an efficient LoRA-based FedFT framework, called LEGEND, by reviewing the inherent characteristics of FedFT.
    \vspace{-0.3cm}
    \item We analyze the joint influence of the LoRA depth and rank distribution on fine-tuning performance and obtain their coupled relationship. 
    Upon this, we develop an efficient algorithm to carefully determine the appropriate LoRA depths with a reasonable rank distribution across all selected LoRA layers for heterogeneous devices to improve the fine-tuning performance.
    \vspace{-0.3cm}
    \item The performance of LEGEND is evaluated through a physical platform with 80 commercial devices that are connected via WiFi.
    The experimental results show that LEGEND provides a speedup of fine-tuning by about 1.5-2.8$\times$ and saves communication costs by about 42.3\% when achieving the target accuracy, compared to existing solutions.
\end{itemize}
\vspace{-0.5cm}

\begin{figure*}[t]
    \centering
    \includegraphics[width=0.99\textwidth]{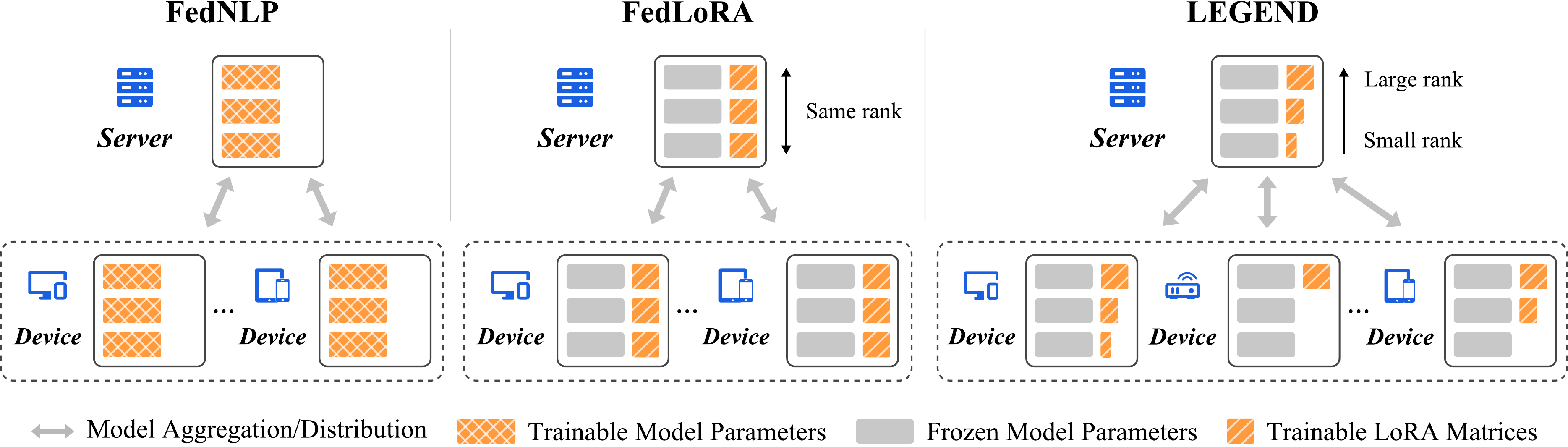}
    \vspace{-0.3cm}
    \caption{Illustration of FedNLP, FedLoRA, and LEGEND. 
    FedNLP (left) fine-tunes all parameters of the LM; FedLoRA (mid) applies the same LoRA configuration to all devices; LEGEND (right) applies different LoRA configurations (\eg, LoRA depth) to devices with heterogeneous capabilities.}
    \label{fig:illustration}
    \vspace{-0.4cm}
\end{figure*}


%% file: contents/prelim.tex
\vspace{-0.2cm}
\subsection{Federated Fine-Tuning LMs with LoRA}
\vspace{-0.1cm}
\textbf{Language Models.}
From design to deployment, training transformer-based LMs typically involves two main stages: pre-training and fine-tuning \cite{vaswani2017attention, lin2021fednlp, cai2022fedadapter}.
In the first stage, the LMs are pre-trained on the large-scale corpus, \eg, Wikipedia \cite{wikidump}, C4 \cite{2019t5}, to learn the ubiquitous linguistic structure, which is independent of downstream tasks \cite{cai2022fedadapter}.
The pre-training stage demands massive computing resources, typically undertaken by large tech companies \cite{devlin2018bert, brown2020language}, such as Google and Microsoft.
Fine-tuning adapts the pre-trained LMs to various downstream tasks, such as text classification and text generation, which require substantial data to update the entire LM for a giving task.
However, fine-tuning the entire LM usually demands excessive resources, \eg, computing and communication resources, leading to slow fine-tuning speeds on resource-constraint devices \cite{cai2022fedadapter, xu2023federated}.

\textbf{LoRA for LMs.}
Herein, low-rank adaptation (LoRA) adds trainable rank decomposition matrices to each transformer layer of the LM while freezing the pre-trained weights of the LM to improve fine-tuning efficiency \cite{hu2021lora}. 
For a pre-trained weight matrix $\mathcal{M} \in \mathbb{R}^{m \times q}$ ($m$ and $q$ are the dimension sizes of $\mathcal{M}$), LoRA injects low-rank decomposition $\Delta \mathcal{M} = \mathcal{B}\mathcal{A}$ as the trainable parameters. 
Note that, $\mathcal{B} \in \mathbb{R}^{m \times r}$ and $\mathcal{A} \in \mathbb{R}^{r \times q}$ are the project-down matrix and the project-up matrix respectively, where $r$ denotes the rank of LoRA and is much smaller than both $m$ and $q$.
Formally, for a simple linear layer $y = \mathcal{M}x$ in the transformer layer, LoRA modifies the forward propagation as: 
\begin{equation}
    y = \mathcal{M}x + \mathcal{B}\mathcal{A}x
    \vspace{-0.1cm}
\end{equation}
where $x \in \mathbb{R}^{q \times s}$ and $y \in \mathbb{R}^{m \times s}$ are the input tensors and the output tensors, respectively, and $s$ is the sequence length (\ie, the number of tokens in a given sequence). 
As a generation of full fine-tuning \cite{hu2021lora}, LoRA utilizes two bypass matrices (\ie, $\mathcal{B}$ and $\mathcal{A}$) with the low rank to perform fine-tuning without modifying the pre-trained weight matrix, thus LoRA can be applied to any transformer layers of the LMs.
Fine-tuning LMs with LoRA can greatly reduce the number of trainable parameters (typically less than 1\% \cite{hu2021lora, zhang2023fedpetuning}) while maintaining satisfactory fine-tuning performance.

\textbf{FedFT with LoRA.}
Considering a distributed system with one PS and $n$ devices, FedFT is proposed to fine-tune the LMs through a loose federation of devices. 
Naturally, LoRA is introduced into FedFT (\eg, FedLoRA) to reduce computing and communication overhead \cite{zhang2023fedpetuning, cho2023heterogeneous}.
The difference between FedFT (\eg, FedNLP) and FedLoRA is illustrated in Figure \ref{fig:illustration}.
FedLoRA only exchanges lightweight LoRA layers $\widetilde{\theta}$ instead of all cumbersome LMs $\overline{\theta}$ in FedFT.
The goal of FedLoRA is to find the optimal model $\theta^* = \{ \widetilde{\theta}^*, \overline{\theta} \}$ minimizing the loss function $f(\theta)$ as follows:
\vspace{-0.2cm}
\begin{equation}
    \min_{\theta = \{ \widetilde{\theta}, \overline{\theta} \}} f(\theta) \triangleq \frac{1}{n} \sum_{i = 1}^{n} f_i(\theta_i)
    \vspace{-0.2cm}
\end{equation}
where $f_i(\theta_i) = \frac{1}{|\mathbb{D}_i|} \sum_{\xi_i \in \mathbb{D}_i} F_i(\theta_i; \xi_i)$ denotes the loss function of the local model $\theta_i$ on device $i$ and $F_i(\theta_i; \xi_i)$ is the loss over data samples $\xi_i$ on local dataset $\mathbb{D}_i$. 
To minimize the local objective function, device $i$ only updates the LoRA layers through the gradient descent algorithm (\eg, AdamW \cite{loshchilov2017decoupled}).
The optimization procedure for updating the LoRA layers $\widetilde{\theta}_i^{h}$ on device $i$ at local step $t$ in round $h$ can be expressed as: 
\begin{equation}
    \widetilde{\theta}_i^{h,t} = \widetilde{\theta}_i^{h,t - 1} - \eta \cdot \nabla f_i(\widetilde{\theta}_i^{h, t - 1})
\end{equation}
where $\eta$ is the learning rate and $\nabla f_i(\widetilde{\theta}_i^{h, t - 1})$ is the gradient of the loss for LoRA layers $\widetilde{\theta}_i^{h, t - 1}$.

After local fine-tuning, all devices send the updated LoRA layers to the PS for global aggregation as follows: 
\vspace{-0.1cm}
\begin{equation}
    \widetilde{\theta}^{h+1} = \frac{1}{n} \sum_{i=1}^{n} \widetilde{\theta}_i^h
    \vspace{-0.1cm}
\end{equation}
After global aggregation, the PS distributes the latest LoRA layers to all devices and moves to the next fine-tuning round.
\vspace{-0.3cm}

\subsection{Importance of LoRA Position}
\begin{figure}[t]
    \centering
    \includegraphics[width=1.0\columnwidth]{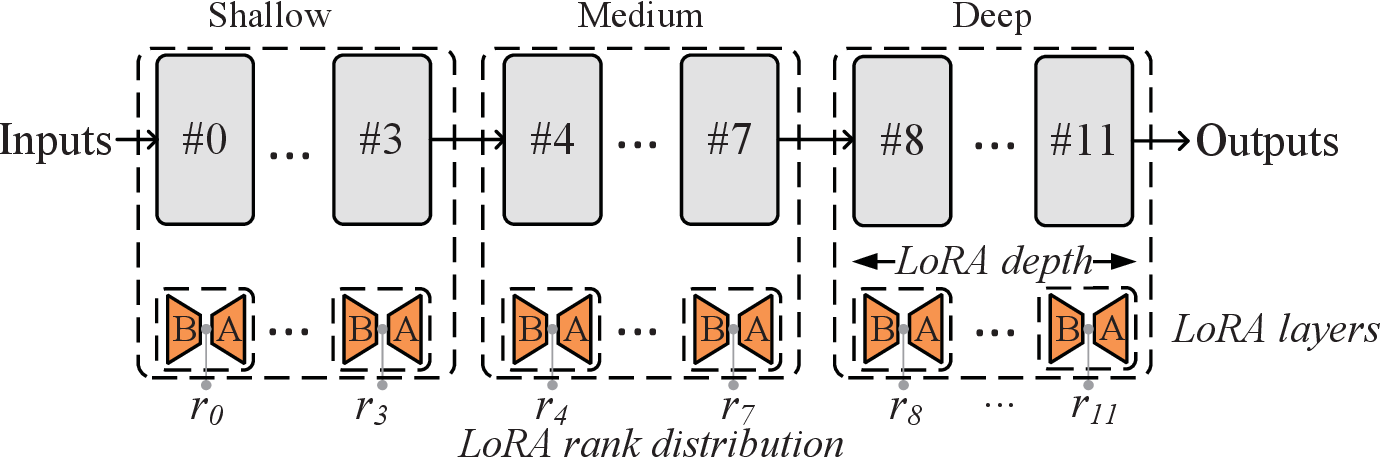}
    \vspace{-0.7cm}
    \caption{Fine-tuning RoBERTa at different positions.
    }
    \label{illu-bert}
    \vspace{-0.4cm}
\end{figure}

\vspace{-0.1cm}
\label{lora-position}
Existing LoRA-based frameworks (\eg, FedLoRA \cite{zhang2023fedpetuning} and HetLoRA \cite{cho2023heterogeneous}) typically add LoRA layers to all transformer layers, which requires the complete backpropagation process to update all LoRA layers \cite{kaddour2023challenges} and results in a slow convergence rate for fine-tuning pre-trained LMs on resource-constrained devices.
As illustrated in Figure \ref{illu-bert}, we divide an LM (\eg, RoBERTa) into three parts representing different positions, \ie, shallow, medium and deep, respectively.
Based on the backward direction of the backpropagation process, fine-tuning the continuous LoRA layers added to the transformer layers at deep position is computationally efficient while achieving satisfactory fine-tuning performance. 
Wei \etal \cite{wei2024flexora} provide rigorous theoretical insights into the convergence of partial LoRA layer fine-tuning.
Besides, only the LoRA layers at deep position need to perform the backpropagation process and be transmitted to the PS, effectively reducing computing/communication overhead and speeding up the fine-tuning process.
In addition, since pre-trained LMs have acquired powerful language understanding and generation capabilities during the pre-training stage, fine-tuning the added LoRA layers to partial transformer layers at deep position can essentially achieve comparable fine-tuning performance \cite{guo2019spottune, lin2021fednlp, zhang2023adaptive}.

\begin{figure}[t]
    \centering
    \subfigure[Test accuracy]{
        \hspace{-5mm}
        \begin{minipage}[t]{0.5\linewidth}
        \centering
        \includegraphics[width=1.6in]{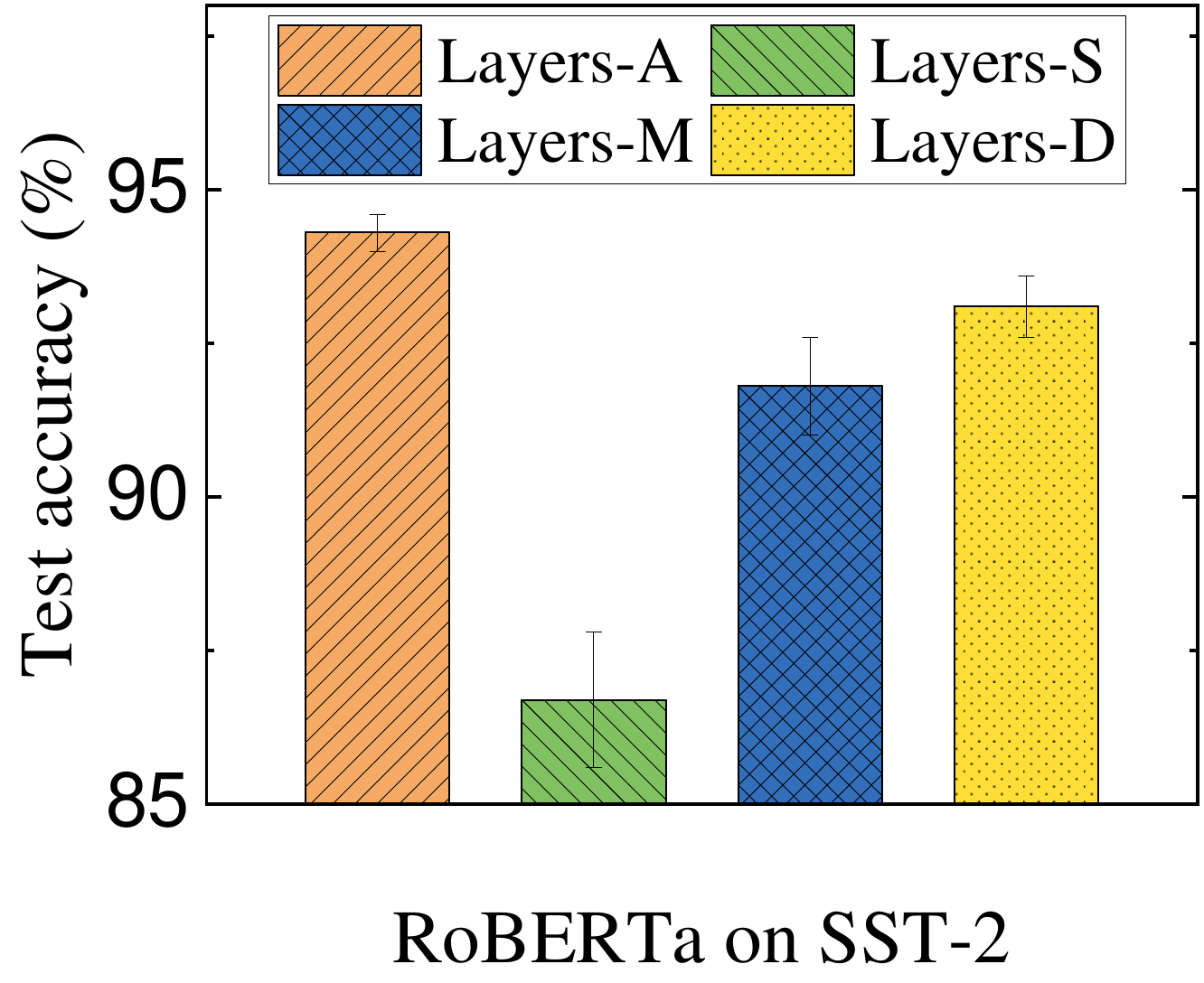}
        \end{minipage}%
        \label{1a}
    }%
    \subfigure[Per-batch latency]{
        \begin{minipage}[t]{0.5\linewidth}
        \centering
        \includegraphics[width=1.6in]{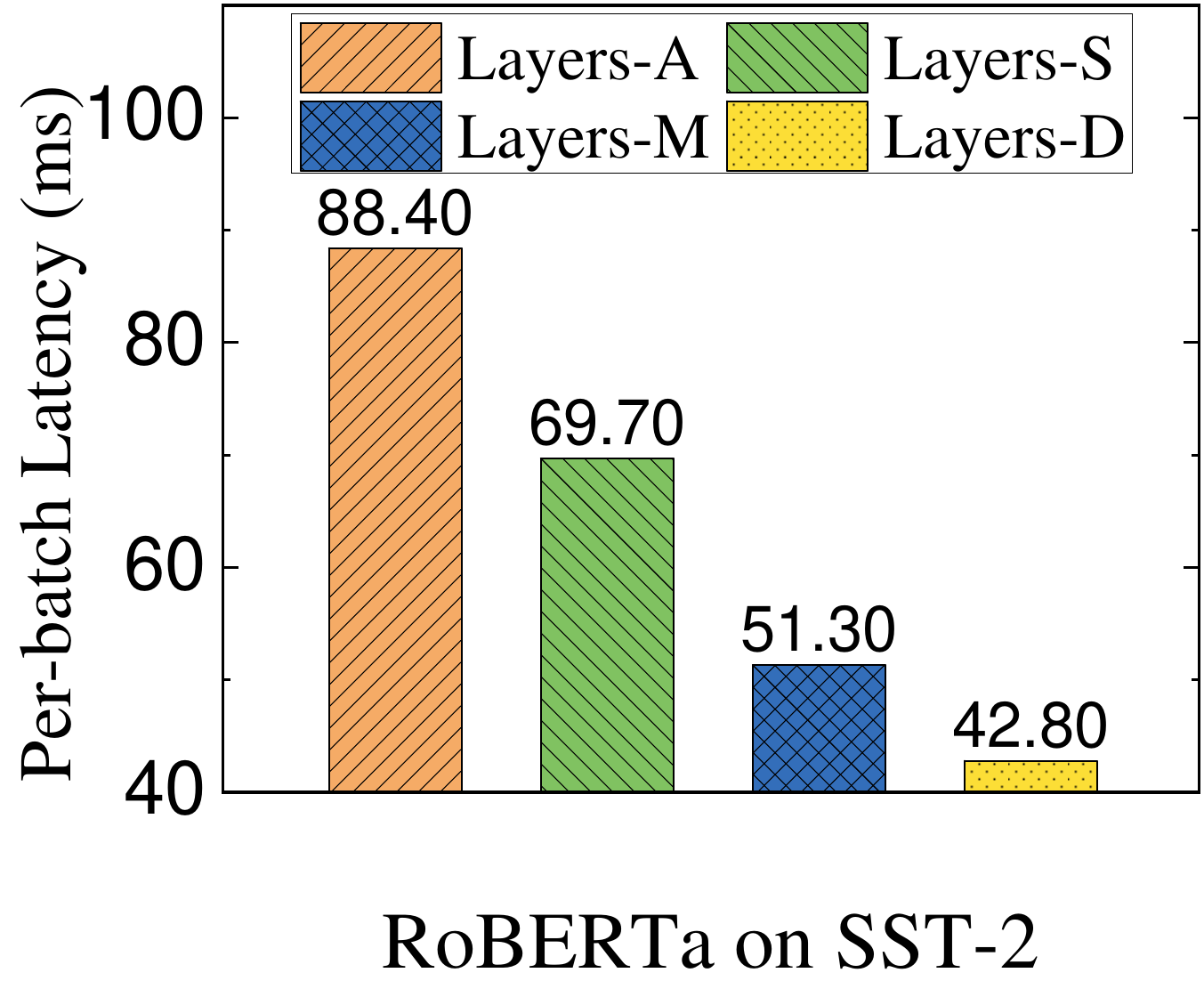}
        \end{minipage}%
        \label{1b}
    }%
    \centering
    \vspace{-0.45cm}
    \caption{The impact of LoRA position.}
    \label{ob-1}
    \vspace{-0.2cm}
\end{figure}

To demonstrate the importance of LoRA position, we conduct a set of experiments for federated fine-tuning RoBERTa \cite{liu2019roberta} on SST-2 \cite{wang2018glue} with 10 devices (more experimental setup details in Section \ref{evaluation}).
We train a 12-layer RoBERTa with LoRA layers added to partial transformer layers at different positions (as illustrated in Figure \ref{illu-bert}), including all layers (denoted as \textit{Layers-A}), shallow layers \{\#0, \#1, \#2, \#3\} (denoted as \textit{Layers-S}), medium layers \{\#4, \#5, \#6, \#7\} (denoted as \textit{Layers-M}), and deep layers \{\#8, \#9, \#10, \#11\} (denoted as \textit{Layers-D}).
As shown in Figure \ref{ob-1}, we can derive the following conclusions:

1) Fine-tuning LoRA layers added to only the transformer layers at the deep position achieves comparable performance to vanilla LoRA.
For example, as shown in Figure \ref{1a}, Layers-D achieves comparable final accuracy (\ie, 93.1\%) to Layers-A (94.3\%), with only fine-tuning a third of LoRA layers in Layers-A (\ie, less computing resource). In addition, compared with Layers-S and Layers-M, Layers-D improves the final model accuracy by 6.4\% and 1.3\%, respectively.

2) Fine-tuning LoRA layers added to only the transformer layers at deep position speeds up the fine-tuning process. 
For instance, by Figure \ref{1b}, Layers-D achieves 2.1$\times$, 1.6$\times$, and 1.2$\times$ speedup compared to Layers-A, Layers-S, and Layers-M, respectively.
This is because Layers-D greatly shortens the backpropagation process from 12 transformer layers to 4 transformer layers, reducing the fine-tuning time.

By the results, the LoRA position will greatly affect fine-tuning performance and resource consumption, while adding LoRA layers to a fixed number of transformer layers at the deep position can save resource cost with satisfactory fine-tuning performance.
Our conclusion is consistent with the relevant conclusions that layers at the deep position are more important than those at the shallow position \cite{zhang2022adaptive, cai2022fedadapter, zhu2023pockengine}.
\vspace{-0.3cm}



\subsection{Importance of LoRA Depth}
\label{lora-depth}
In addition to the results of adding LoRA layers to partial transformer layers at deep position, we further explore the impact of the number of transformer layers at deep position to add extra LoRA layers (called LoRA depth) on fine-tuning performance.
In general, LoRA depth is closely related to fine-tuning performance and resource consumption.
Specifically, the larger LoRA depth means fine-tuning more layers at deep positions to enhance fine-tuning performance but leads to a longer backpropagation process (\ie, slower convergence rate).
Although the smaller LoRA depth can shorten the backpropagation process to speed up the fine-tuning process, it constrains the number of tunable transformer layers and thus restricts the task-fitting capability of the LM.
Therefore, it is critical and non-negligible to determine the appropriate LoRA depth for achieving the trade-off between fine-tuning performance and resource consumption.

\begin{figure}[t]
    \centering
    \subfigure[Accuracy loss \& Per-batch latency]{
        \begin{minipage}[t]{0.5\linewidth}
        \centering
        \includegraphics[width=1.6in]{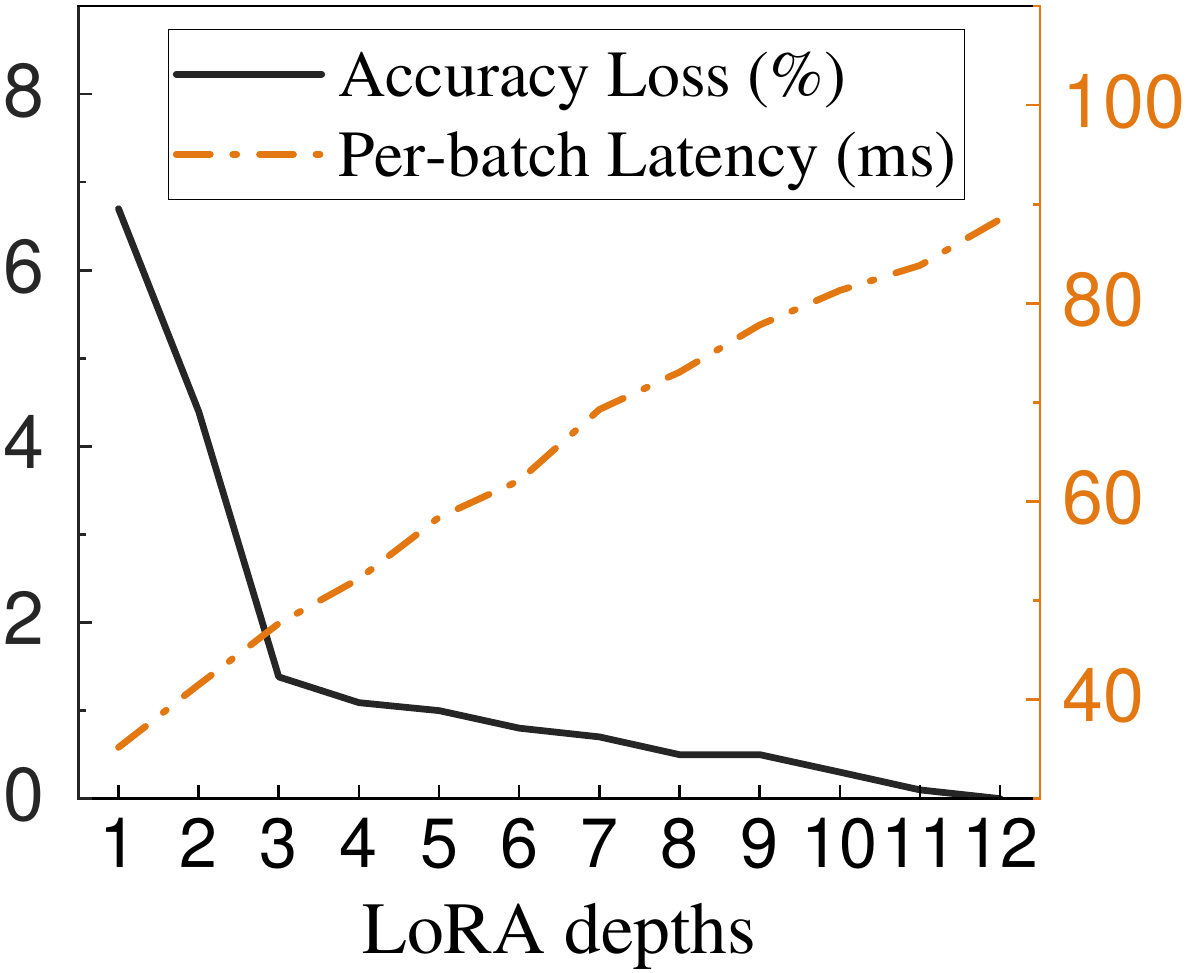}
        \end{minipage}%
        \label{2a}
    }%
    \subfigure[Memory usage]{
        \hspace{-5mm}
        \begin{minipage}[t]{0.5\linewidth}
        \centering
        \includegraphics[width=1.6in]{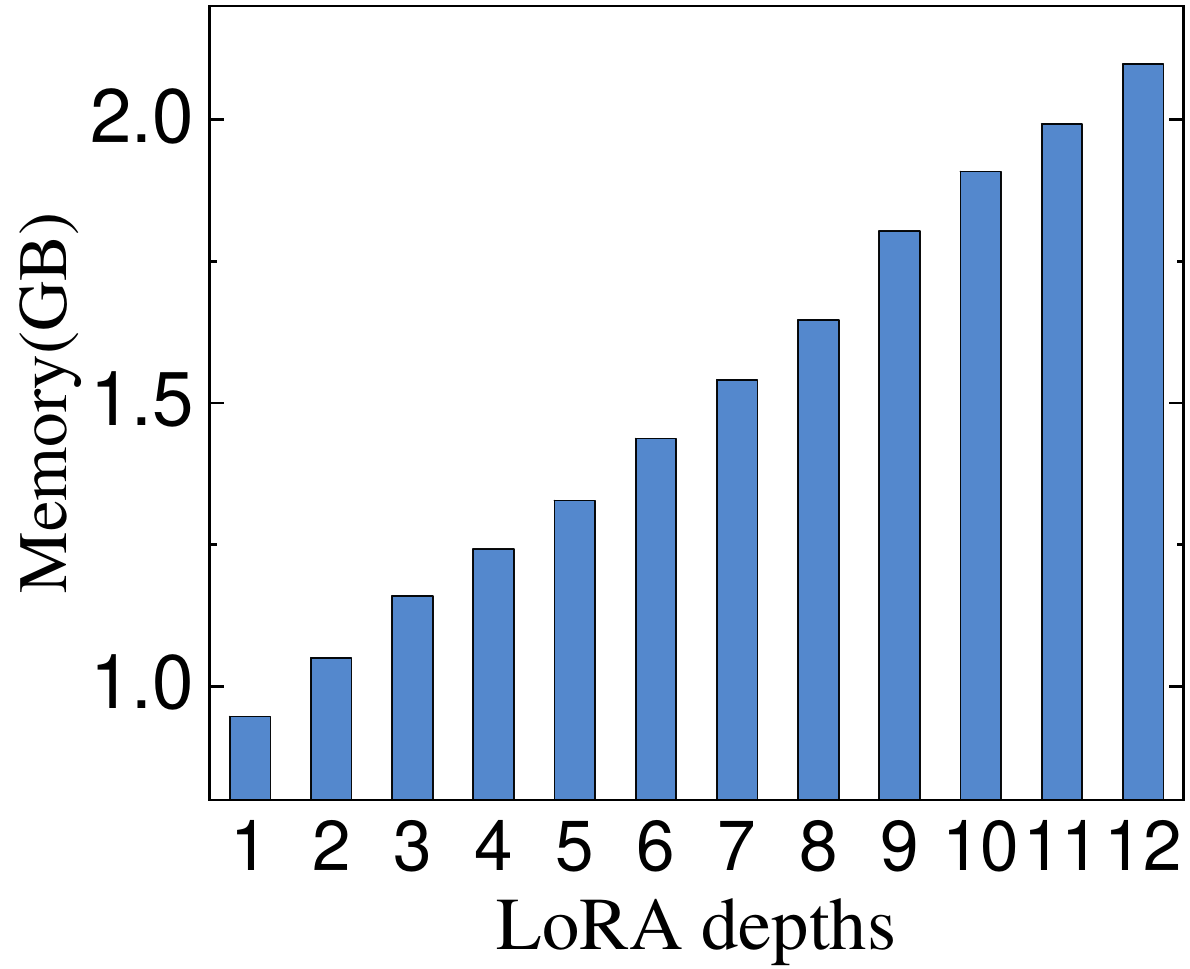}
        \end{minipage}%
        \label{2b}
    }%
    \centering
    \vspace{-0.5cm}
    \caption{The impact of LoRA depth.}
    \label{ob-2}
    \vspace{-0.4cm}
\end{figure}

To illustrate the impact of the LoRA depth, we conduct a set of experiments for fine-tuning RoBERTa on SST-2 with different LoRA depths (from 1 to 12) and rank of 8. 
We can draw the following conclusions from Figure \ref{ob-2}:

1) As LoRA depth increases, resource consumption (\eg, memory usage) increases almost linearly, resulting in slow convergence rate.
For example, by Figures \ref{2a} and \ref{2b}, with each additional LoRA layer, the per-batch latency increases by approximately 5ms, and the memory usage increases by approximately 107MB. 
Compared with LoRA depth of 1, fine-tuning RoBERTa with depth of 12 results in a 252\% increase in per-batch latency and 221\% growth in memory usage. 

2) The fine-tuning performance improves with the increase of LoRA depth, but the magnitude of the improvement diminishes gradually.
For instance, by Figure \ref{2a}, compared to the LoRA depth of 3, fine-tuning RoBERTa with LoRA depths of 6 and 9 exhibit improvements of 0.6\% and 0.9\% in accuracy, respectively.
As LoRA depth increases from 1 to 3, the final model performance improves by 5.3\%, but only 1.4\% from LoRA depth of 3 to 12.

By the results, carefully determining LoRA depth is critical to improve fine-tuning performance while saving computing and communication resources.
Moreover, to deal with system heterogeneity, LEGEND assigns different LoRA depths for heterogeneous devices to adapt their capabilities.
The devices with strong computing and communication capabilities are assigned with larger LoRA depths, while those with lower computing and communication capabilities are assigned with smaller LoRA depths, so as to reduce waiting time and further improve fine-tuning efficiency.
\vspace{-0.3cm}

\begin{figure}[t]
    \hspace{-3mm}
    \centering
    \subfigure[Performance gain (rank 1 $\rightarrow$ 128)]{
        \begin{minipage}[t]{0.5\linewidth}
        \centering
        \includegraphics[width=1.62in]{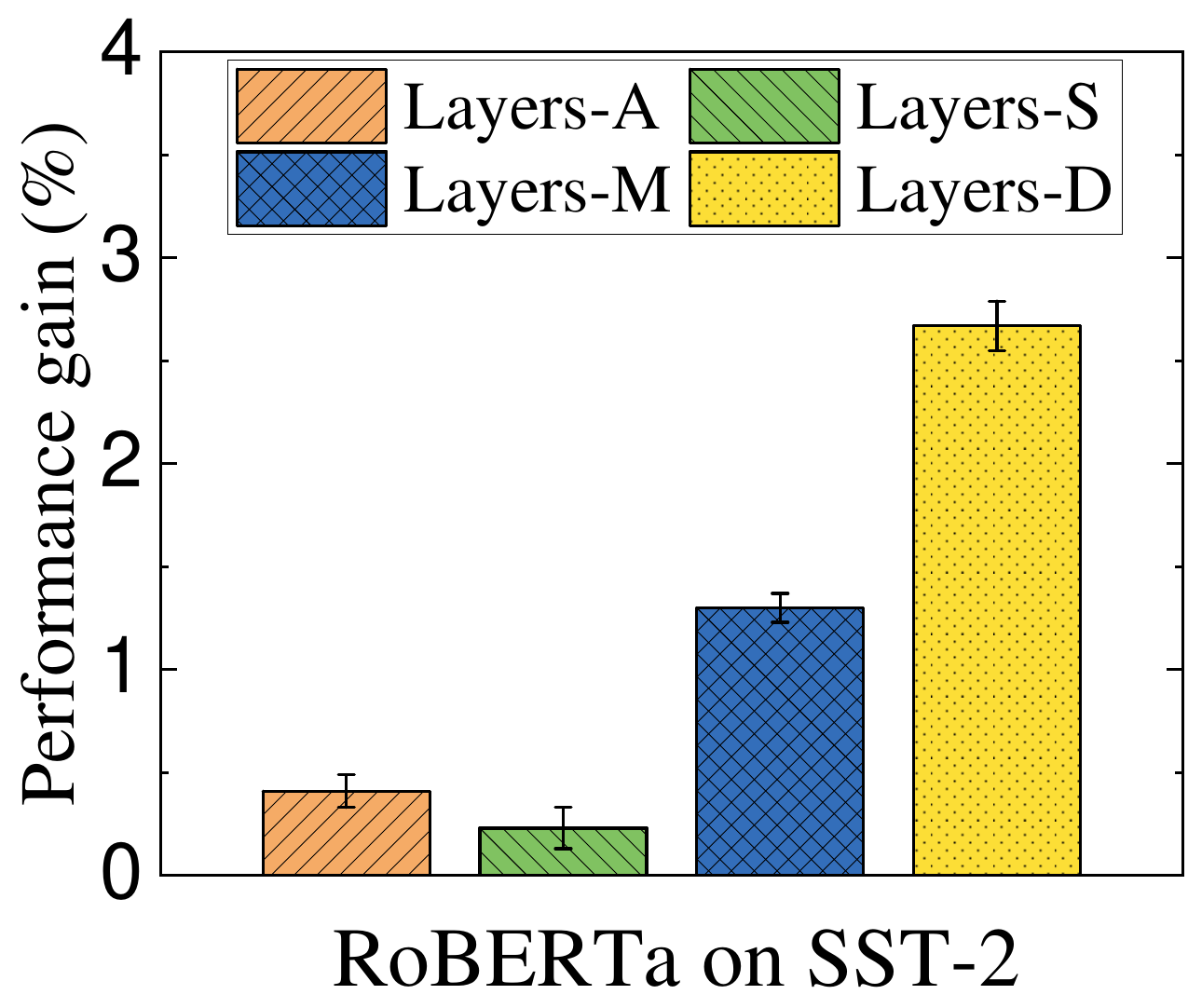}
        \end{minipage}%
        \label{3a}
    }%
    \subfigure[Final accuracy]{
        \begin{minipage}[t]{0.5\linewidth}
        \centering
        \includegraphics[width=1.7in]{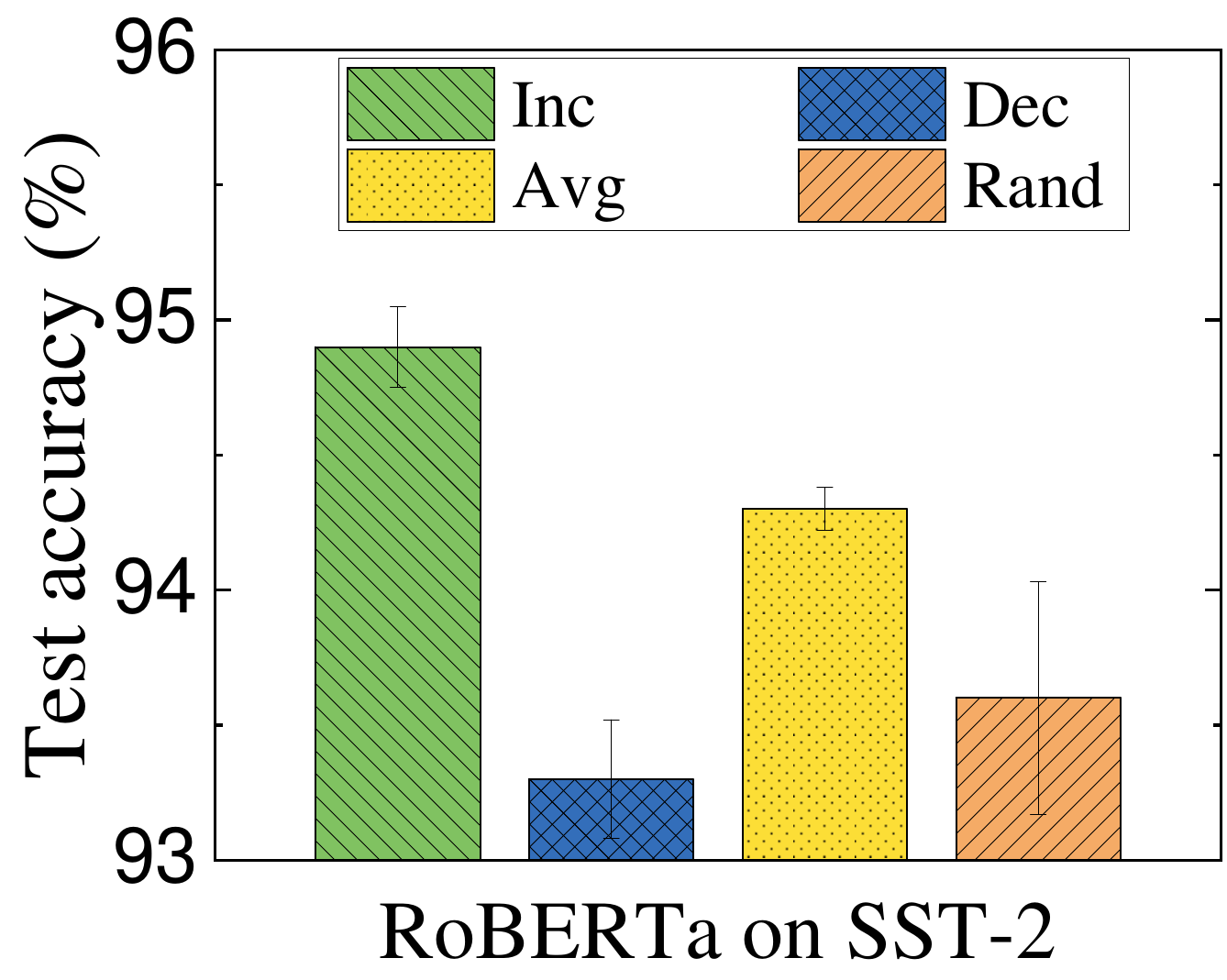}
        \end{minipage}%
        \label{3b}
    }%
    \centering
    \vspace{-0.5cm}
    \caption{The impact of LoRA rank distribution.}
    \label{ob-3}
    \vspace{-0.4cm}
\end{figure}

\subsection{Importance of LoRA Rank Distribution}
\label{lora-distribution}
We further explore the impact of rank distribution on fine-tuning performance. 
In general, the rank of LoRA layer is closely related to the model capacity, and increasing rank is essential for better performance \cite{song2024increasing, shuttleworth2024lora}. 
Given a fixed total rank budget under resource constraints, strategically allocating higher ranks to task-relevant layers becomes crucial for maximizing model performance \cite{zhang2022adaptive, cho2023heterogeneous}.
Specifically, the higher ranks of the deep LoRA layers achieve better fine-tuning performance on a specific task, since the layers at deep position need to focus on higher semantic levels and more contextual information \cite{lin2021fednlp, ma2022layer, liu2023yoga, gao2024higher}. 
Shallow LoRA layers with smaller ranks leverage the pre-trained model's inherent feature extraction capability, preventing overfitting while maintaining model performance. 

To demonstrate the impact of rank distribution, we conduct two sets of experiments for fine-tuning RoBERTa on SST-2. 
Firstly, we separately apply the different ranks, \ie, \{1, 2, 4, 8, 16, 32, 64, 128\}, for Layers-A, Layers-S, Layers-M, and Layers-D (refer to Section \ref{lora-position}) and record their performance gains for comparison.
Secondly, we employ four different rank distributions with a total budget of 96 across 12 transformer layers (from input to output): Inc [4, 4, 5, 6, 7, 7, 8, 9, 10, 11, 12, 13], Dec [13, 12, 11, 10, 9, 8, 7, 7, 6, 5, 4, 4], Avg [8, 8, 8, 8, 8, 8, 8, 8, 8, 8, 8, 8], and Rand (randomly allocated).
The findings illustrated in Figure \ref{ob-3} highlight that:
%

1) Deeper layers are more sensitive to the rank of LoRA.  
Specifically, applying larger ranks for LoRA layers close to the output helps to achieve better model performance.
For instance, as shown in Figure \ref{3a}, the final performance gain of Layers-A, Layers-S, Layers-M, and Layers-D are separately 0.41\%, 0.23\%, 2.67\%, and 1.3\%.

2) The gradually increasing rank distribution achieves higher model accuracy. 
For example, as illustrated in Figure \ref{3b}, Inc distribution achieves the best accuracy (94.9\%) among all variants, outperforming others by up to 1.6\%. 
This is because Inc strategically allocates larger ranks to critical deeper layers, greatly enhancing fine-tuning performance.


Based on the results, it is effective and necessary to allocate larger ranks to deeper layers. 
However, due to the limited computing and communication resources, allocating larger ranks to deeper layers constrains the applicable LoRA depth, whereas smaller ranks tend to slow down the convergence process.
Therefore, in this paper, LEGEND simultaneously determines the appropriate LoRA depth with reasonable rank distribution for different devices so as to balance the trade-off between resource consumption and fine-tuning performance.
\vspace{-0.2cm}

\begin{figure}[t]
    \centering
    \includegraphics[width=1\columnwidth]{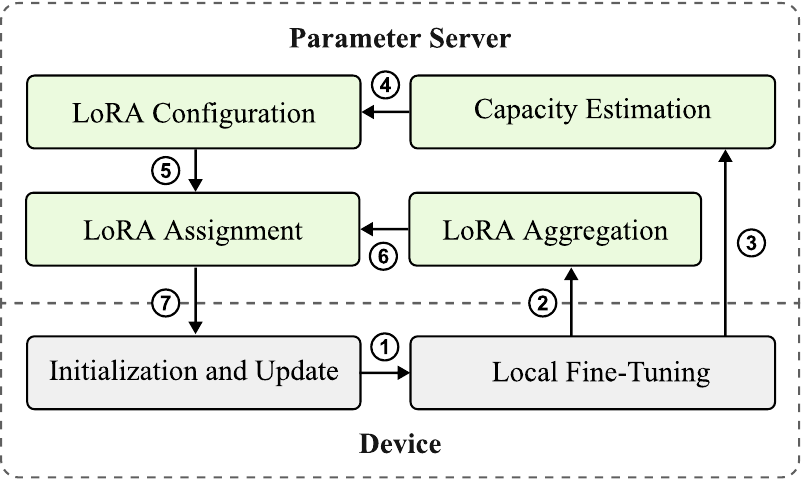}
    \vspace{-0.5cm}
    \caption{The proposed LEGEND framework. 
    }
    \label{framework}
\vspace{-0.4cm}
\end{figure}

%% file: contents/overview.tex
\vspace{-0.2cm}
As illustrated in Figure \ref{framework}, LEGEND consists of two key components with a total of six main modules, \ie,  four modules on the PS and two modules on each device. 
The details of each module are as follows:

\textbf{Initialization and Update.} 
The PS distributes different LoRA layers for heterogeneous devices to adapt their capabilities.
Hence, the devices need to initialize and update the local model (\normalsize{\textcircled{\scriptsize{1}}}) for local fine-tuning in each round.

\textbf{Local Fine-Tuning.}
Based on the initialized local model, the devices perform local fine-tuning and record the fine-tuning status information (\eg, computing time and communication time).
After that, the devices send the updated LoRA layers (\normalsize{\textcircled{\scriptsize{2}}}) and the status information (\normalsize{\textcircled{\scriptsize{3}}}) to the PS.

\textbf{Capacity Estimation.}
To make effective LoRA configurations, the PS estimates the capabilities of each device (\normalsize{\textcircled{\scriptsize{4}}}) by calculating the moving average of the historical status information of the devices.

\textbf{LoRA Configuration.}
In this module, the PS simultaneously determines the appropriate LoRA depth with rank distribution, \ie, LoRA configurations (\normalsize{\textcircled{\scriptsize{5}}}), for the devices.

\textbf{LoRA Aggregation.}
The PS performs adaptive weighted aggregation for the collected LoRA layers with different LoRA depths from all devices to obtain the aggregated global LoRA layers (\normalsize{\textcircled{\scriptsize{6}}}).

\textbf{LoRA Assignment.}
Due to the diverse LoRA configurations of devices, the PS needs to assign the specific LoRA layers (\normalsize{\textcircled{\scriptsize{7}}}) to each device based on the aggregated model.
\vspace{-0.4cm}



%% file: contents/design.tex
\vspace{-0.3cm}
\subsection{\bluenote{Initialization and Update}}
\vspace{-0.2cm}
In each round, the device receives a set of LoRA layers from the PS to initialize and update the local model for local fine-tuning. 
These LoRA layers are tailored to the device's capacity (\eg, computing and communication capacity), including information on which transformer layers to fine-tune and the ranks of those LoRA layers. 
Specifically, device $i$ receives the LoRA layers $\widetilde{\theta}_i^h = \{ \widetilde{\theta}_{i, l}^h | l \in [L - {k}_i^h, L - 1] \}$ to initialize and update ${k}_i^h$ transformer layers close to the output, where $L$ denotes the number of transformer layers in the pre-trained model $\overline{\theta}$. 
The LoRA layers $\widetilde{\theta}_{i, l}^h \in \widetilde{\theta}_i^h$ includes coupled LoRA matrices for all linear layers in transformer layer $l$.
The coupled LoRA matrices include two components, \ie, the project-down matrix $\mathcal{B} \in \mathbb{R}^{m \times r}$ and the project-up matrix $\mathcal{A} \in \mathbb{R}^{r \times q}$, where $r$ denotes the rank of the LoRA matrix and is much smaller than both $m$ and $q$.
Without loss of generality, for arbitary coupled LoRA matrices $\mathcal{B}_{i,l}^{h} \in \widetilde{\theta}_{i,l}^h$ and $\mathcal{A}_{i,l}^{h} \in \widetilde{\theta}_{i,l}^h$, the device injects the matrices into the linear layer $y = M_l \cdot x$ with pre-trained weight matrix $M_l$ in transformer layer $l$ as $\{ ( \mathcal{B}_{i,l}^{h}, \mathcal{A}_{i,l}^{h}, M_l ) | l \in \mathcal{L}_i^h \}$.
Then, the forward propagation of the linear layer with initialized parameters $\{ \mathcal{B}_{i,l}^{h}, \mathcal{A}_{i,l}^{h}, M_l\}$ in $l$-th transformer layer on device $i$ in round $h$ can be expressed as: 
\vspace{-0.2cm}
\begin{equation}
    y = M_l \cdot x + \mathcal{B}_{i,l}^{h} \cdot \mathcal{A}_{i,l}^{h} \cdot x
    \vspace{-0.2cm}
\end{equation}
where $x$ is the given input tensor and $y$ is the respective output tensor of the initialized linear layer. 
After that, the device initializes and updates the local model $\theta_i^h = \{ \widetilde{\theta}_i^h, \overline{\theta} \}$ for local fine-tuning in round $h$. 
\vspace{-0.2cm}

\subsection{Local Fine-Tuning} 
\vspace{-0.1cm}
After local model initialization, device $i$ fine-tunes the model on local dataset $\mathbb{D}_i$.
During the process of local fine-tuning in round $h$, device $i$ is associated with the local loss function $f_i(\theta_i^h)$, where $\theta_i^h = \{ \widetilde{\theta}_i^h, \overline{\theta} \}$ is the local model.
The loss of device $i$ on the local dataset $\mathbb{D}_i$ in round $h$ are as follows:
\vspace{-0.3cm}
\begin{equation}
    f_i(\theta_i^h) = \frac{1}{|\mathbb{D}_i|} \sum_{\xi \in \mathbb{D}_i} F_i(\theta_i^h; \xi_i)
    \vspace{-0.3cm}
\end{equation}
where $\xi_i$ is a batch of data samples in $\mathbb{D}_i$, and $F_i(\theta_i^h; \xi_i)$ is the local loss over $\xi_i$. 
In general, the devices leverage stochastic gradient descent, \eg, AdamW \cite{loshchilov2017decoupled}, to iteratively update the LoRA layers based on the computed gradients over each batch of data samples in $\mathbb{D}_i$ \cite{lin2021fednlp, cai2022fedadapter}.
Specifically, for a batch of local data samples $\xi_i$ on device $i$, the process of updating the LoRA layers $\widetilde{\theta}_i^{h}$ on device $i$ at local step $t$ in round $h$ is expressed as: 
\vspace{-0.2cm}
\begin{equation}
    \widetilde{\theta}_i^{h,t} = \widetilde{\theta}_i^{h,t - 1} - \eta \cdot \nabla f_i(\widetilde{\theta}_i^{h, t - 1})
\end{equation}
where $\eta$ is the learning rate and $\nabla f_i(\widetilde{\theta}_i^{h, t - 1})$ is the gradient of the loss for LoRA layers $\widetilde{\theta}_i^{h, t - 1}$.
When completing local fine-tuning, the devices send the updated local LoRA layers to the PS for aggregation.
Simultaneously, the devices upload the relevant computing and communication information of the current round to PS for device resource estimation.
\vspace{-0.2cm}

\subsection{Capacity Estimation}
\vspace{-0.1cm}
The estimation of device capabilities (\eg, time-varying computing and communication capabilities) is essential for LEGEND to determine reasonable LoRA configurations for heterogeneous devices. 
For arbitrary device $i$ in round $h$, LEGEND utilizes $\mu_i^h$ to denote the time required for updating all LoRA layers in a transformer layer during backpropagation, which can be recorded by the devices directly during the process of local fine-tuning, to indicate the computing capability. 
Besides, since the upload bandwidth is usually much smaller than the download bandwidth in typical WANs \cite{mcmahan2017communication, konecny2016federated} and the size of LoRA layers is typically less than 1\% of the original model size \cite{zhang2023fedpetuning}, LEGEND employs the uploading time $\beta_i^h$ of transmitting a LoRA layer with unit rank from the device $i$ in round $h$ to the PS to indicate the communication capability.

In round $h$, PS collects recent computing time $\hat{\mu}_i^h$ and uploading time $\hat{\beta}_i^h$ from device $i$ and maintains the historical status.
Then, we use the moving average with the historical status of devices to estimate the capacities of the devices \cite{leroy2019federated}. 
Accordingly, the PS estimates the computing time $\mu_i^h$ and the uploading time $\beta_i^h$ for device $i$ in round $h$ by calculating the moving average with $\rho \in [0, 1]$ (for example, $\rho = 0.8$ in our experiments) as:
\vspace{-0.1cm}
\begin{equation}
\mu_i^h = \rho \cdot \mu_i^{h - 1} + (1 - \rho) \cdot \hat{\mu}_i^h, \forall i \in [1,n], \forall h \in [1,H]
\vspace{-0.1cm}
\end{equation}
\begin{equation}
\ \ \ \ \beta_i^h = \rho \cdot \beta_i^{h - 1} + (1 - \rho) \cdot \hat{\beta}_i^h, \forall i \in [1,n], \forall h \in [1,H]
\vspace{-0.1cm}
\end{equation}
The primary focus of this work is not on improving status estimation techniques and advanced methods \cite{halperin2010predictable, yue2017linkforecast} can be easily integrated into LEGEND.
\vspace{-0.2cm}

\subsection{LoRA Configuration}
\vspace{-0.1cm}
LEGEND addresses the challenges of resource constraints and system heterogeneity by determining appropriate LoRA configurations, \ie, LoRA depth and rank distribution, for heterogeneous devices to promote fine-tuning efficiency. 
Based on the observation in Sections \ref{lora-depth} and \ref{lora-distribution}, LEGEND adaptively assigns LoRA depth for device $i$ in round $h$ with gradually increasing rank distribution.
The LoRA configuration of device $i$ in round $h$ can be denoted as the rank distribution $R_i^h = \{ r_{i,l} | l \in [L - k_i^h, L - 1]\}$ that includes the LoRA depth $k_i^h$, where $r_{i,l}$ is the rank of the LoRA layers in $l$-th transformer layer.
For simplicity, we use $\mathcal{L}_i^h = [L - k_i^h, L - 1]$ to represent the index of the deep $k_i^h$ transformer layers. 
Assuming that the total rank of all $L$ transformer layers is $\psi$, the constraints of rank distribution for device $i$ in round $h$ are as:
\vspace{-0.2cm}
\begin{equation}
    r_{i,l - 1} \leq r_{i,l}
    \vspace{-0.1cm}
\end{equation}
\begin{equation}
    \sum_{l \in \mathcal{L}_i^h} r_{i,l} \leq \psi
    \vspace{-0.1cm}
\end{equation}

In round $h$, based on the estimation of computing and communication capacities, the completion time (including computing and communication time) on device $i$ is expressed as:
\vspace{-0.3cm}
\begin{equation}
    t_i^h =  \hat{t_i} + k_i^h \cdot \mu_{i}^h + \sum_{l \in \mathcal{L}_i^h} r_{i,l} \cdot \beta_i^h
\end{equation}
where $\hat{t_i}$ represents the computing time of forward propagation in one round of local fine-tuning, $k_i^h \cdot \mu_{i}^h$ represents the total time for backpropagation during the process of local fine-tuning, and $\sum_{l \in \mathcal{L}_i^h} r_{i,l} \cdot \beta_i^h$ denotes the uploading time.
Additionally, the waiting time for device $i$ can be represented as $t^h - t_i^h$, where $t^h = \max\{ t_i^h| i \in [1, n] \}$ denotes the completion time of the slowest device in round $h$. 
Then, the average waiting time of all devices in round $h$ can be formulated as:
\begin{equation}
    \mathcal{W}^h = \frac{1}{n} \sum_{i = 1}^n (t^h - t_i^h)
    \vspace{-0.3cm}
\end{equation}

We define the computing resource consumption of the LoRA layers in a transformer layer with unit rank during local fine-tuning as $c$.
In addition, we use $\hat{c}$ to denote the computing resource consumption for the pre-trained model $\overline{\theta}$ during the forward propagation, which is a constant.
Then, assuming that the total computing resource budgets of device $i$ in round $h$ is $C_i^h$, the computing resource constraints can be expressed as follows:
\vspace{-0.2cm}
\begin{equation}
    \label{constrain-comp}
     \hat{c} + \sum_{l \in \mathcal{L}_i^h} r_{i,l} \cdot c \leq C_i^h
     \vspace{-0.3cm}
\end{equation}
Similarly, we use $b$ to denote the communication consumption of LoRA layers in a transformer layer with unit rank.
Let $B_i^h$ represent the total communication resource budget for device $i$ in round $h$. 
Then, the communication resource constraints can then be formulated as:
\vspace{-0.2cm}
\begin{equation}
    \label{constrain-comm}
    \sum_{l \in \mathcal{L}_i^h} r_{i,l} \cdot b \leq B_i^h
    \vspace{-0.3cm}
\end{equation}

Given a specific task in the FedFT system, LEGEND determines an appropriate LoRA configuration $R_i^h$ for device $i$ according to the estimation of the device's resource in round $h$ to minimize the overall fine-tuning time $\sum_{h = 1}^H t^h$. 
Thus, we can formulate the problem as follows:

\centerline{$\min \sum\limits_{h=1}^{H} t^h$}
\vspace{-0.5cm}
\begin{equation}\label{problem}
    s.t.
    \begin{cases}
    r_{i,l - 1} \leq r_{i,l}, & \forall l \in \mathcal{L}_i^h\\
    \sum_{l \in \mathcal{L}_i^h} r_{i,l} \leq \psi, & \forall i \in [1, n] \\
    t_i^h =  \hat{t_i} + k_i^h \cdot \mu_{i}^h + \sum_{l \in \mathcal{L}_i^h} r_{i,l} \cdot \beta_i^h, & \forall i \in [1, n]\\
    \mathcal{W}^h = \frac{1}{n} \sum_{i=1}^n (t^h - t_i^h) \leq \epsilon, & \forall h \in [1, H]\\
    \hat{c} + \sum_{l \in \mathcal{L}_i^h} r_{i,l} \cdot c \leq C_i^h, & \forall i \in [1, n] \vspace{0.5ex} \\
    \sum_{l \in \mathcal{L}_i^h} r_{i,l} \cdot \hat{b} \leq B_i^h, & \forall i \in [1,n] \vspace{0.5ex} \\
    \end{cases}
    \vspace{-0.1cm}
\end{equation}
\vspace{-0.4cm}

The first set of inequalities denotes the constraints of rank distribution.
The second set of inequalities represents that the rank of all LoRA layers cannot exceed the given budget $\psi$. 
The third set of equations represents the total time of local fine-tuning and uploading on device $i$ in round $h$.
The fourth set of inequalities represents the average waiting time of all devices cannot exceed the predefined threshold $\epsilon > 0$.
The last two sets of inequalities express that the accumulated computing and communication resources cost cannot exceed the computing and communication resource budgets of device $i$ in round $h$, respectively.

\begin{algorithm}[t]
\caption{ LoRA Configuration Determination}\label{alg:lora-config}
\label{alg:device-group}
\begin{algorithmic}[1]
\Require Total rank budget $\psi$; the completion time of the devices $t_i^h$; parameter $\lambda$.
\Ensure LoRA configuration $R_i^h = \{ r_{i,l}|l \in [L - k_i^h, L - 1] \}$.
\Function{LoraConfiguration}{\ }: 
\State Calculate the gap between the maximum and minimum LoRA depth in round $h$
            \vspace{-0.4cm}
            \begin{equation}
                k^h \leftarrow \lceil L \cdot \frac{t^h - t_{i,\min}^h\}}{t^h} \rceil; \nonumber
                \vspace{-0.4cm}
            \end{equation}
\State Determine the LoRA depth for device $i$ by
    \vspace{-0.4cm}
    \begin{equation}
        \vspace{-0.4cm}
        k_i^h \leftarrow \lceil k^h \cdot \frac{t^h - t_i^h}{t^h} \rceil; \nonumber
    \end{equation}
\State Get global rank distribution $R = \{r_l | l \in [0, L - 1]\}$ using an arithmetic sequence with a common difference of $\lambda$, where $r_l = r_{l - 1} + \lambda$.
\State Adjust LoRA depth $k_i^h$ to ensure the configuration meets device-specific computing and communication constraints by Equations \ref{constrain-comp} and \ref{constrain-comm}.
\State Generate LoRA configuration $R_i^h = \{ r_{i,l}^h| l \in [L - k_i^h, L - 1] \}$ for device $i$, where $r_{i,l}^h = r_l \in R$.
\EndFunction
\end{algorithmic}
\end{algorithm}

To solve the problem in Equation (\ref{problem}), we propose a greedy-based LoRA configuration determination algorithm, termed LCD, as shown in Algorithm \ref{alg:lora-config}. 
First, the PS calculates the gap $k^h$ between the maximum and minimum LoRA depth (Line 2) according to the device status information. 
Then, the completion time gap $k_i^h$ between device $i$ and the slowest is used to generate the appropriate LoRA depth to adapt heterogeneous devices (Line 3). 
For example, the most powerful device is assigned with maximum LoRA depth $L$, the weakest device is assigned with LoRA depth of $L - k^h$, while other devices are assigned values within the gap $[L - k^h, L]$ based on their capabilities.
Secondly, the PS gets the reasonable rank distribution $R = \{r_l | l \in [0, L - 1]$ for all transformer layers using an arithmetic sequence with a common difference of $\lambda$ in round $h$ (line 4). 
We use the parameter $\lambda$ to guide the generation of the rank distribution, where $\lambda$ indicates that the rank of LoRA layers added to the transformer layer $l$ is $\lambda$ greater than that of the adjacent transformer layer $l-1$, \ie, $r_l = r_{l - 1} + \lambda$.
For instance, $\lambda = 1$ in our experiments by default. 
Then, we greedily adjust the LoRA depth to ensure the configuration meets the device's capacity (Line 5).
Finally, based on global rank distribution $R$ and LoRA depth $k_i^h$, the PS generates the LoRA configuration $R_i^h = \{ r_{i,l}^h| l \in [L - k_i^h, L - 1] \}$ for device $i$ in round $h$ (Line 6).
\vspace{-0.4cm}

\subsection{LoRA Aggregation}
\vspace{-0.1cm}
After receiving the updated LoRA layers from all devices, the PS performs global aggregation.
Since the LoRA depth varies across devices while the rank of each LoRA layer remains consistent across all devices, the PS performs adaptive layer-wise aggregation on the collected LoRA layers, \ie, aggregating each layer based on the number of devices contributing to that layer.
We use $\widetilde{\theta}^{h+1}_l$ to represent the global LoRA layers in $l$-th transformer layer obtained by aggregating the respective LoRA layers from $n_l$ devices.
Formally, the adaptive layer-wise aggregation of LoRA layers in transformer layer $l$ can be expressed as follows:
\vspace{-0.3cm}
\begin{equation}
    \widetilde{\theta}^{h+1}_l = \frac{1}{n_l} \sum_{i = 1}^{n_l} \widetilde{\theta}_{i,l}^{h}
    \vspace{-0.3cm}
\end{equation}
After aggregation, PS obtains a set of global LoRA layers $\widetilde{\theta}^{h+1} = \{ \widetilde{\theta}^{h+1}_l | l \in [0, L - 1] \}$, which will be used for assigning different LoRA layers for each device.

\subsection{LoRA Assignment}
\vspace{-0.2cm}
According to the obtained LoRA configuration, the PS assigns specific LoRA layers for each device.
Specifically, based on the LoRA configuration $R_i^h = \{r_{i,l} | l \in \mathcal{L}_i^h\}$ and the aggregated set of global LoRA layers that can be expressed as:
\vspace{-0.1cm}
\begin{equation}
    \widetilde{\theta}^{h} = \{ \widetilde{\theta}_{l}^{h} | l \in [0, L - 1] \}
    \vspace{-0.1cm}
\end{equation}
LEGEND generates LoRA layers $\widetilde{\theta}_i^h$ for device $i$ by selecting the LoRA layers from global LoRA layers $\widetilde{\theta}^{h}$ as follows:
\vspace{-0.1cm}
\begin{equation}
    \widetilde{\theta}_i^{h} = \{ \widetilde{\theta}_{i, l}^{h} | l \in \mathcal{L}_i^h\}
    \vspace{-0.1cm}
\end{equation}
where $\widetilde{\theta}_{i, l}^{h} = \widetilde{\theta}_l^{h} \in \widetilde{\theta}^{h}$.
Since the LoRA layers are all continuous layers at deep position, there is no need to send a separate LoRA configuration for local initialization and update.
Thus, after obtaining the LoRA layers, the PS immediately distributes the LoRA layers to the respective devices. 
\vspace{-0.5cm}

%% file: contents/implementation.tex
\vspace{-0.2cm}
\bluenote{
We implement our FedFT prototype based on the open-source FedPETuning framework \cite{zhang2022federated}, extending its functionality with approximately 2.1K lines of custom code. 
The prototype is designed to support heterogeneous computing platforms, with a specific focus on commercial NVIDIA Jetson devices \cite{mittal2019survey}, including Jetson TX2, Jetson NX, and Jetson AGX.
The software platform is built based on Docker Swarm \cite{merkel2014docker, naik2016building}, a distributed software development kit that helps build distributed systems with the ability to monitor the status of each device.
For computing, we utilize PyTorch \cite{paszke2019pytorch} to facilitate the implementation of model fine-tuning on devices, ensuring platform independence while allowing platform-specific backend acceleration. 
To accelerate the on-device fine-tuning, we utilize NVIDIA-provided development packages\footnote{https://forums.developer.nvidia.com/t/pytorch-for-jetson/72048} to take full advantage of the underlying hardware capabilities.
For communication, we adopt MPI (Message Passing Interface) \cite{gabriel2004open}, which includes a collection of sending and receiving functions, \eg, \textit{comm.send(data, dest, tag)}/\textit{comm.recv(sour, tag)}, to streamline communication between the PS and devices.
The prototype provides simple APIs to abstract away the problem of federated fine-tuning on heterogeneous computing platforms.
For example, in the PS, we use docker API \textit{docker stack deploy} to deploy the fine-tuning process on both PS and the heterogeneous device with the customized container for different devices.
The implementation addresses key challenges in supporting heterogeneous devices by developing flexible communication and fine-tuning protocols that can adapt to varying computational resources.
}
\vspace{-0.2cm}


%% file: contents/simulation.tex
\begin{table}[!t]
    \caption{Technical Overview of Jetson Platforms}
    \centering
    \begin{tabular}{lcc}
        \Xhline{1pt}
        \textbf{Jetson} & \textbf{AI Performance} & \textbf{GPU Type} \\ 
        \Xhline{0.7pt}
        TX2 & 1.33 TFLOPS & 256-core Pascal \\ \hline
        NX & 21 TOPS & 384-core Volta\\ \hline
        AGX Xavier & 22 TOPS &   512-core Volta \\ \hline
        \Xhline{1pt}
        \textbf{Jetson} & \textbf{CPU Type} & \textbf{ROM} \\ 
        \Xhline{0.7pt}
        TX2 & Denver 2 and ARM 4 & 8 GB LPDDR4\\ \hline
        NX & 6-core Carmel ARM 8 & 8 GB LPDDR4x\\ \hline
        AGX Xavier& 8-core Carmel ARM 8 & 32 GB LPDDR4x \\ \hline
        \Xhline{1pt}
    \end{tabular}
    \label{jetson-info}
    \vspace{-0.5cm}
\end{table}

\vspace{-0.2cm}
\subsection{Methodology}
\vspace{-0.1cm}
\label{evaluation}
\textbf{Experimental Setup.}
Extensive experiments are conducted on the implemented prototype system with one PS and 80 devices to evaluate the performance of LEGEND.
Specifically, the PS runs on a workstation equipped with an Intel(R) Xeon(R) Platinum 8358P CPU (@ 2.60GHz with 128 cores), 8 NVIDIA RTX A6000 GPUs (48GB memory each) and 512 GB RAM.
In addition, we specify 80 NVIDIA commercial developer kits, including 30 Jetson TX2 kits, 40 Jetson NX kits, and 10 Jetson AGX kits, as devices to construct a heterogeneous system. 
The detailed technical specifications of Jetson TX2, NX, and AGX kits are listed in Table \ref{jetson-info}.

\textbf{Settings of System Heterogeneity.}
To emulate the heterogeneous computing and communication capabilities among devices, we present the following setups.

(1) \textbf{\textit{{For computing.}}}
By specifying different modes of the Jetson devices (\ie, Jetson TX2, NX, and AGX), our prototype system enables these devices to work with varying computing capabilities.
Specifically, Jetson TX2 offers four configurable modes, whereas the Jetson NX and AGX support up to eight modes.
For example, the Jetson AGX with the mode 0 (\ie, the highest performance mode of Jetson AGX) achieves fine-tuning by 100$\times$ faster than the TX2 with the mode 1 (\ie, the lowest performance mode of Jetson TX2).
Besides, to reflect resource varying over time, the devices are configured to randomly change the mode every 20 rounds.

(2) \textbf{\textit{For communication.}}
To replicate the practical network environment, all devices are connected to the PS via Wi-Fi routers in the prototype system.
Concretely, the devices are randomly shuffled and divided into four groups, with each group containing 20 devices. 
Then, these groups are placed at different locations, \ie, 2m, 8m, 14m, and 20m, away from the Wi-Fi routers.
Due to random channel noise and competition among devices, the bandwidth between the PS and devices varies dynamically during the fine-tuning.
The bandwidth of devices is measured by iperf3 \cite{tirumala1999iperf}, which fluctuates between 1Mb/s and 30Mb/s.

\begin{table}[t]
\renewcommand{\arraystretch}{1.1}
    \centering
    \caption{Overview of Datasets for Experimental Evaluation}
    \begin{tabular}{lccccccc}
         \Xhline{1pt}
         \textbf{Dataset} &  \textbf{Partition Rules} & \makecell{\textbf{\# Training}\\\textbf{Samples}} &\makecell{\textbf{\# Test}\\\textbf{Samples}} \\
         \Xhline{0.7pt}
         SST-2  & non-i.i.d. & 67,349   & 1,821  \\ \hline
         QNLI   & non-i.i.d. & 104,743 & 5,463  \\ \hline
         QQP  & non-i.i.d. & 363,846  & 40,430  \\ \hline
         MNLI  & non-i.i.d. & 392,702  & 9,815  \\ \hline
         GSM-8K   & i.i.d. & 7473 & 1,319  \\ \hline
         MMLU  & i.i.d. & 20,000  & 2,000 \\ \hline
         \Xhline{1pt}
    \end{tabular}
    \label{tab:datasets}
    \vspace{-0.5cm}
\end{table}

\textbf{Tasks and Models.} 
\rednote{
We evaluate the performance of LEGEND using three representative models downloaded from Hugging Face \cite{wolf2020transformers}, \ie, RoBERTa \cite{liu2019roberta}, DeBERTa \cite{he2021debertav3}, and Llama \cite{touvron2023llama}, across the three categories of tasks, including general language understanding \cite{wang2018glue}, massive multitask understanding \cite{hendrycks2020measuring} and mathematical reasoning \cite{wang2017deep, cobbe2021training}.
}

\bluenote{
1) \textit{\textbf{General Language Understanding}} aims to assess the natural language understanding capabilities of the models. 
We pick four datasets from the General Language Understanding Evaluation (GLUE) datasets \cite{wang2018glue}, encompassing a diverse spectrum of natural language understanding challenges: SST-2 for sentiment analysis, QNLI for question-based natural language inference, QQP for semantic equivalence, and MNLI for multi-genre textual entailment, respectively.
Following the \textit{Dirichlet} distribution with $\alpha = 10$, we build the non-independent and identically (non-i.i.d.) distributed datasets \cite{lin2021fednlp}. 
We fine-tune a 125M-parameter RoBERTa-base model \cite{liu2019roberta} with 12 transformer layers on SST-2 and QNLI, and a 350M-parameter DeBERTa-Large model \cite{he2021debertav3} with 24 transformer layers on QQP and MNLI, respectively.
}

\begin{figure*}
	\centering
	\subfigure[SST-2]{
		\begin{minipage}[b]{0.23\textwidth}
			\includegraphics[width=1\textwidth]{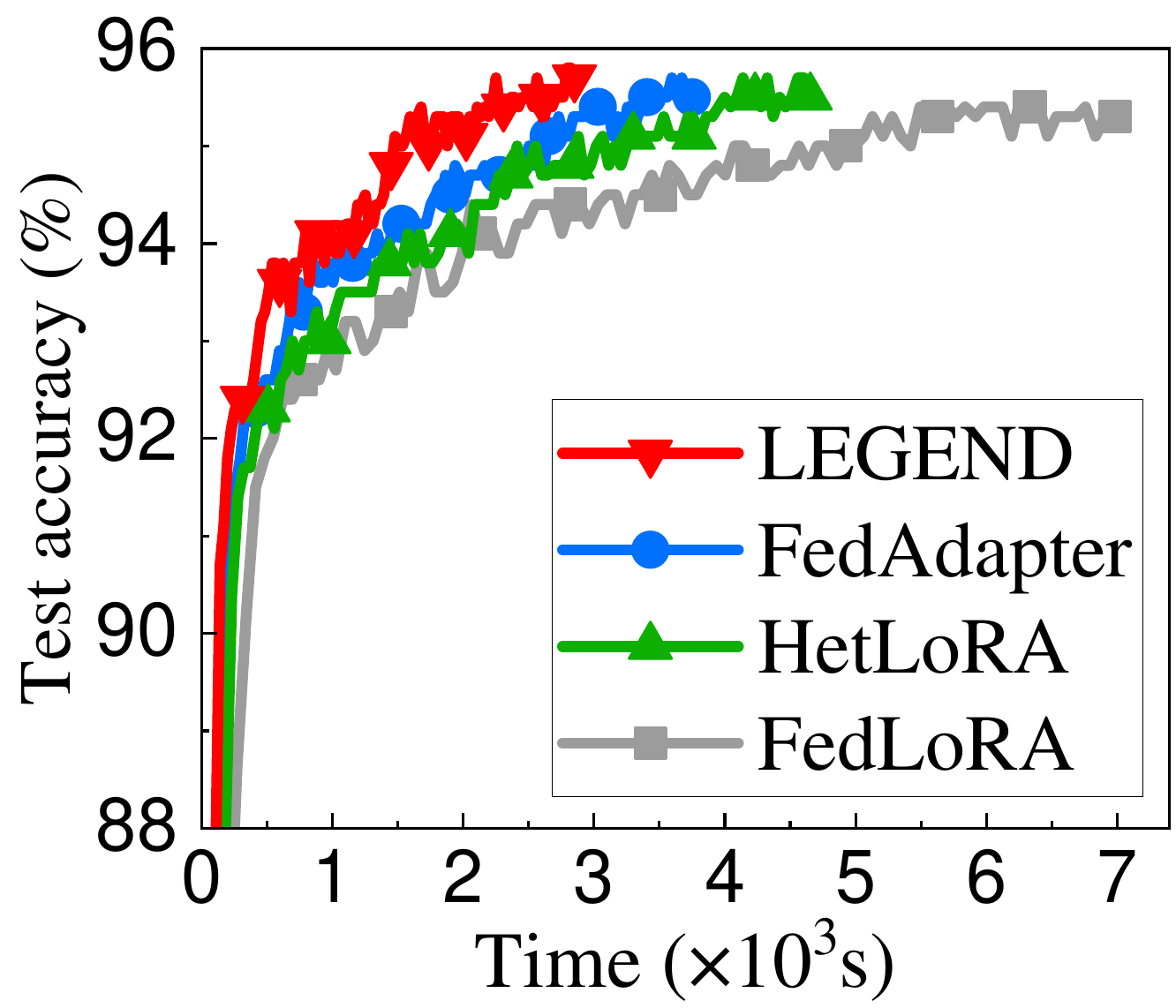} 
		\end{minipage}
		\label{roberta-time2acc-sst2}
	}
    	\subfigure[QNLI]{
    		\begin{minipage}[b]{0.23\textwidth}
   		 	\includegraphics[width=1\textwidth]{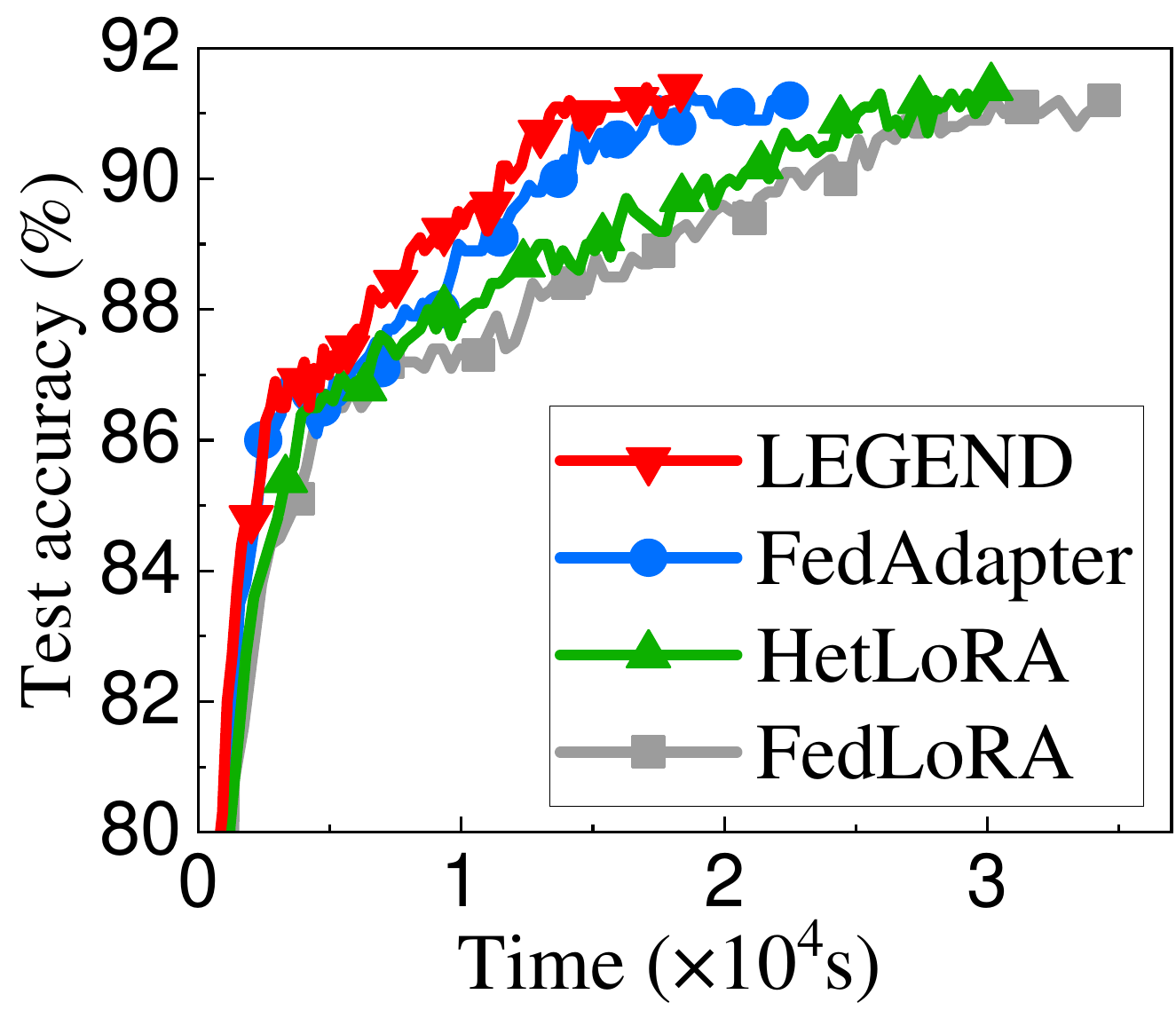}
    		\end{minipage}
		\label{roberta-time2acc-qnli}
    	}
	\subfigure[QQP]{
		\begin{minipage}[b]{0.23\textwidth}
			\includegraphics[width=1\textwidth]{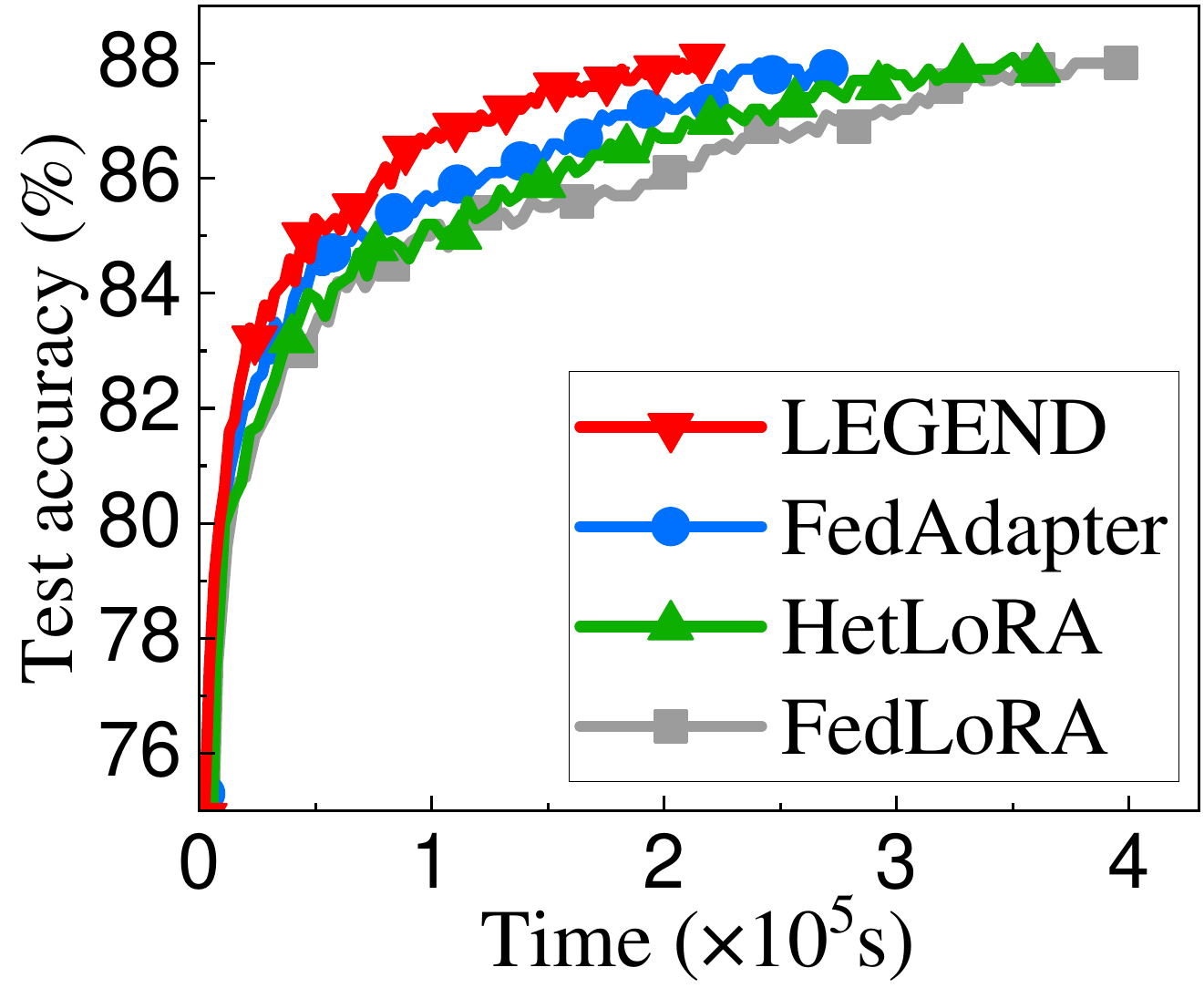} 
		\end{minipage}
		\label{deberta-time2acc-qqp}
	}
    	\subfigure[MNLI]{
    		\begin{minipage}[b]{0.23\textwidth}
		 	\includegraphics[width=1\textwidth]{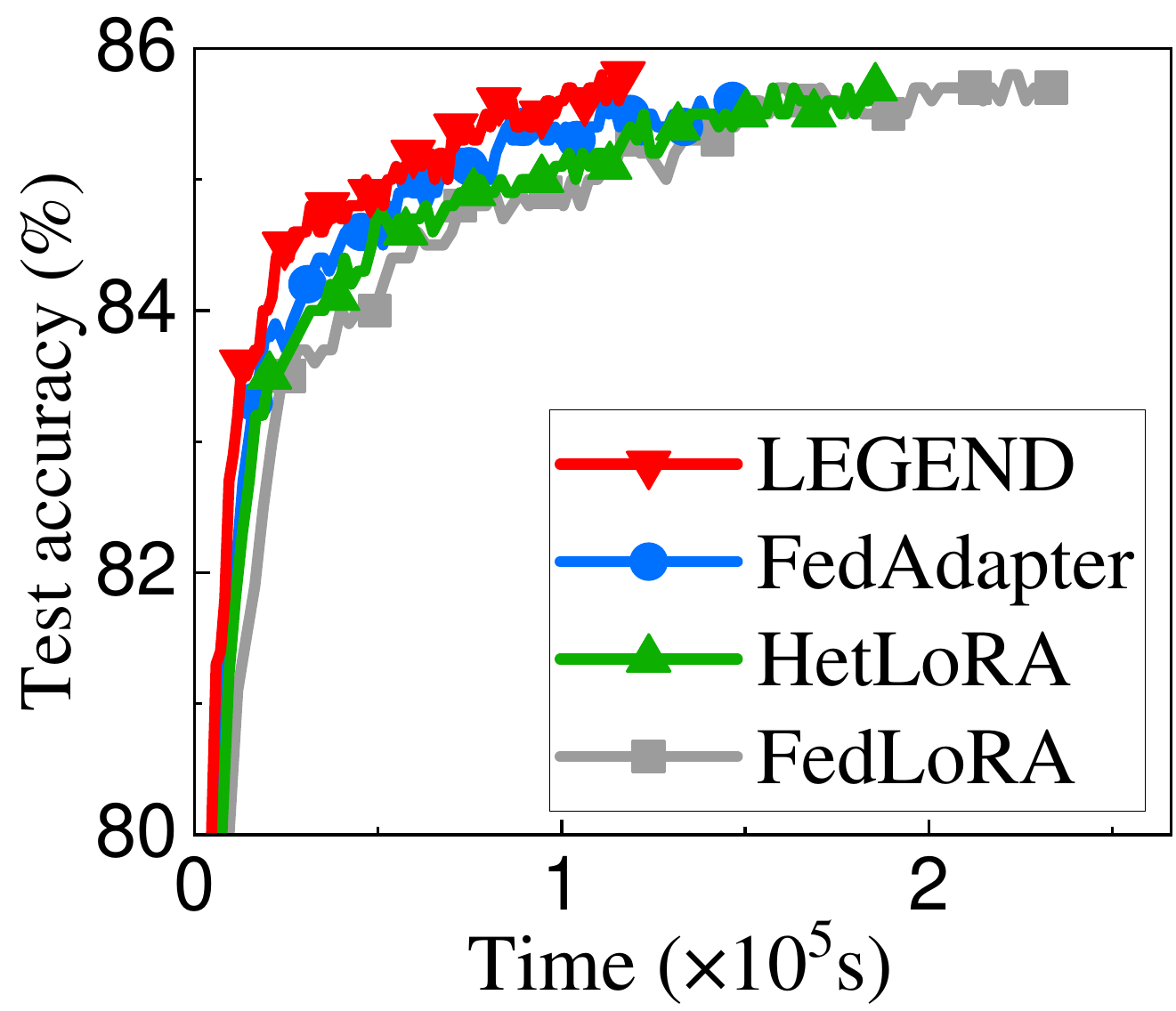}
    		\end{minipage}
		\label{deberta-time2acc-mnli}
    	}
        \vspace{-0.3cm}
	\caption{Fine-tuning process of four approaches on general language understanding tasks.}
        \vspace{-0.4cm}
        \label{time-acc}
\end{figure*}

\begin{figure*}
	\centering
	\subfigure[Time to reach 95\% accuracy]{
		\begin{minipage}[b]{0.23\textwidth}
			\includegraphics[width=1\textwidth]{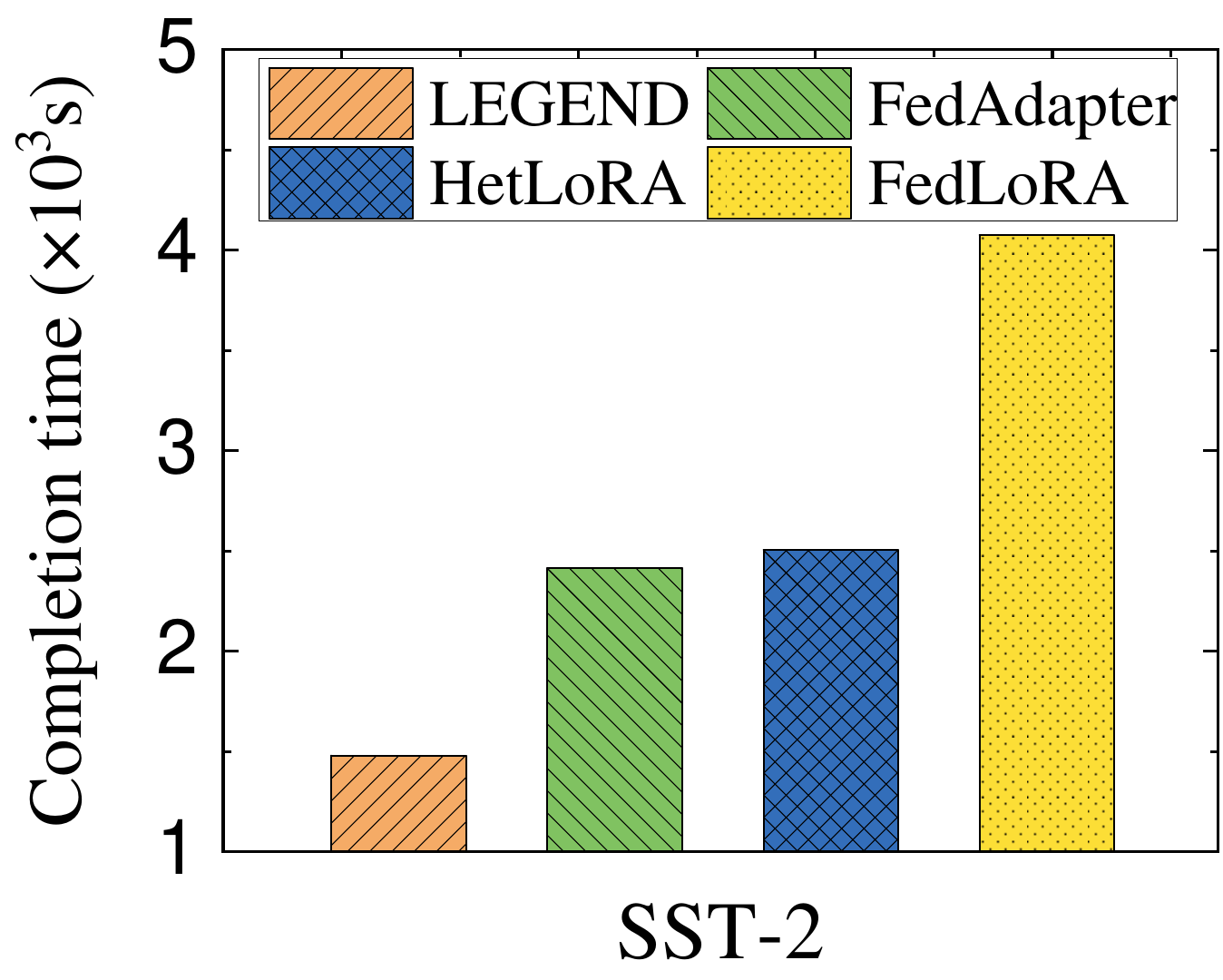} 
		\end{minipage}
		\label{roberta-completion-sst2}
	}
    	\subfigure[Time to reach 90\% accuracy]{
    		\begin{minipage}[b]{0.23\textwidth}
   		 	\includegraphics[width=1\textwidth]{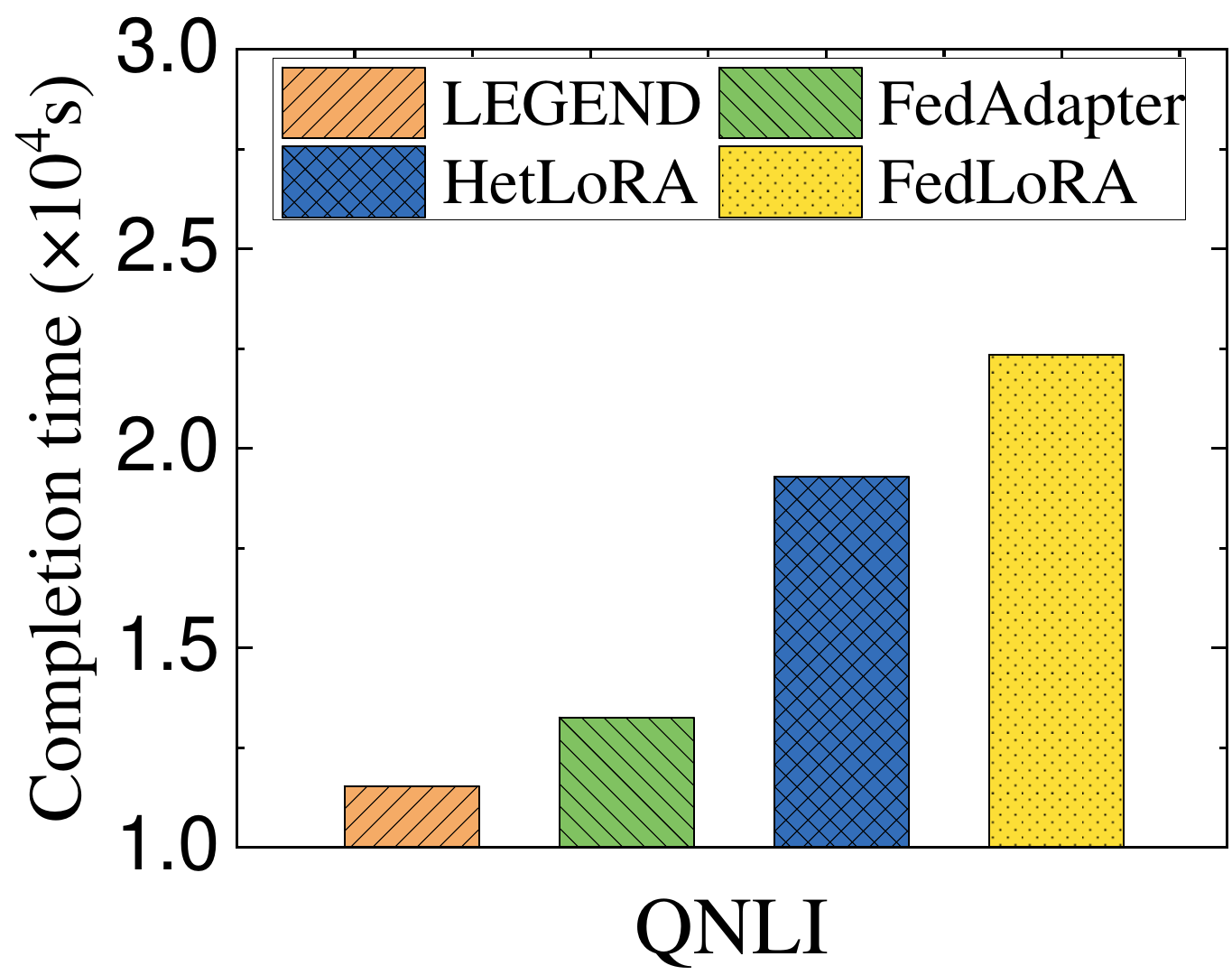}
    		\end{minipage}
		\label{roberta-completion-qnli}
    	}
	\subfigure[Time to reach 87\% accuracy]{
		\begin{minipage}[b]{0.23\textwidth}
			\includegraphics[width=1\textwidth]{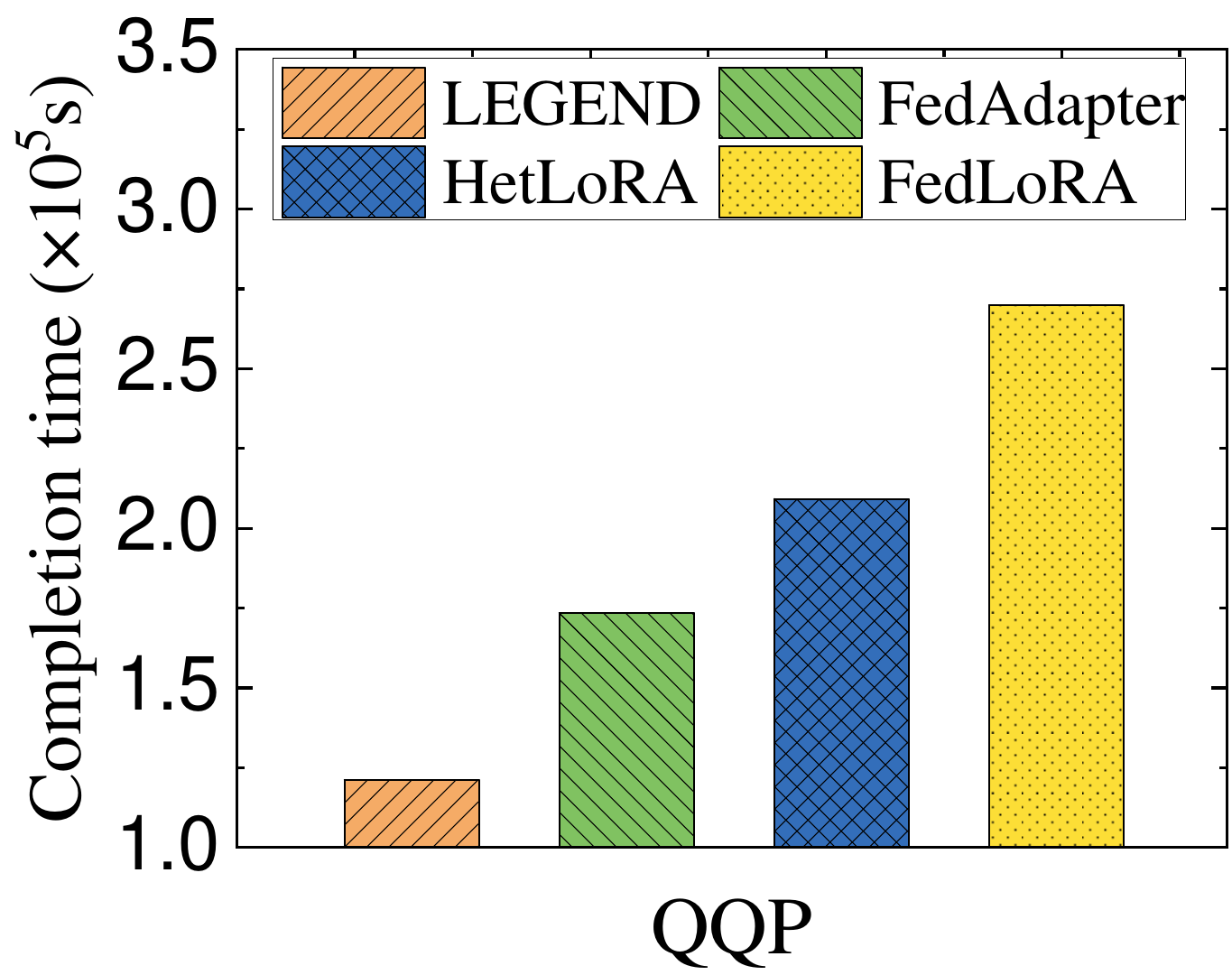} 
		\end{minipage}
		\label{deberta-completion-qqp}
	}
    	\subfigure[Time to reach 85\% accuracy)]{
    		\begin{minipage}[b]{0.23\textwidth}
		 	\includegraphics[width=1\textwidth]{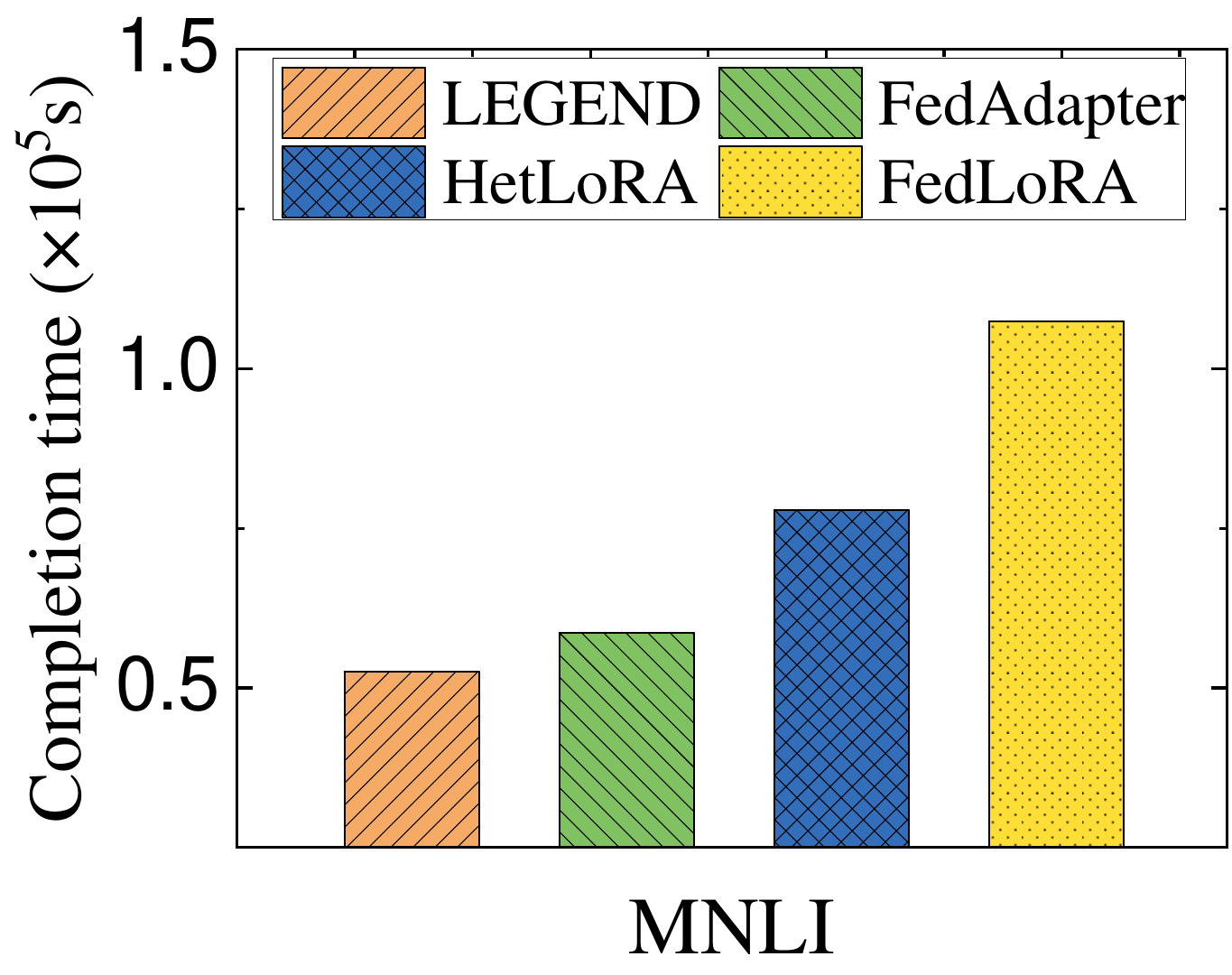}
    		\end{minipage}
		\label{deberta-completion-mnli}
    	}
        \vspace{-0.3cm}
	\caption{Completion time of four approaches on general language understanding tasks.}
        \vspace{-0.4cm}
        \label{completion-time}
\end{figure*}

\begin{figure}[t]
    \hspace{10mm}
    \centering
    \subfigure[{Fine-tuning process (MMLU)}]{
        \hspace{0.7mm}
        \begin{minipage}[b]{0.5\linewidth}
        \centering
        \includegraphics[width=1.6in]{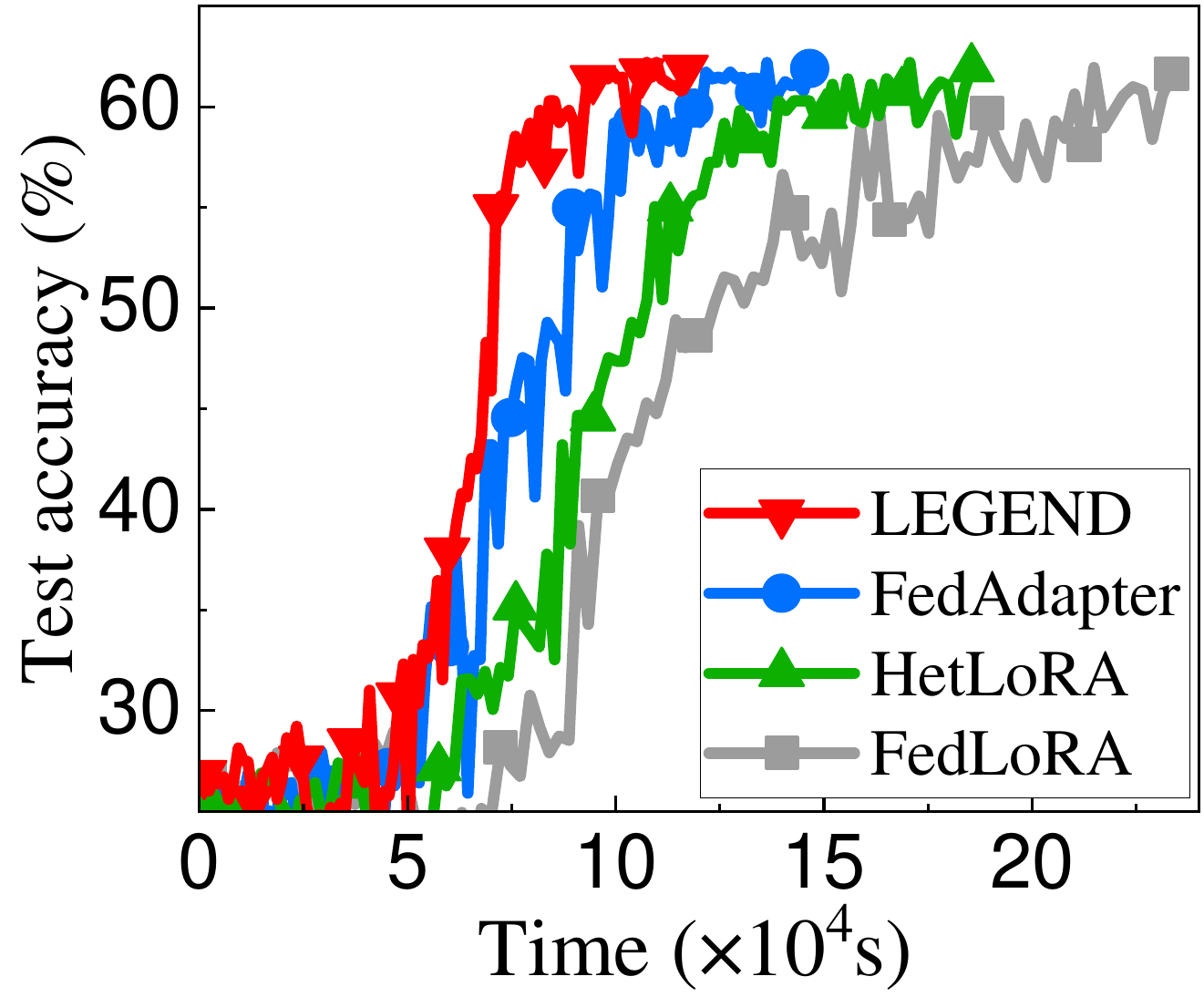}
        \end{minipage}%
        \label{llama2-7b-mmlu-time2acc}
    }%
    \subfigure[{Time to reach 60\% accuracy}]{
        \begin{minipage}[b]{0.5\linewidth}
        \centering
        \includegraphics[width=1.6in]{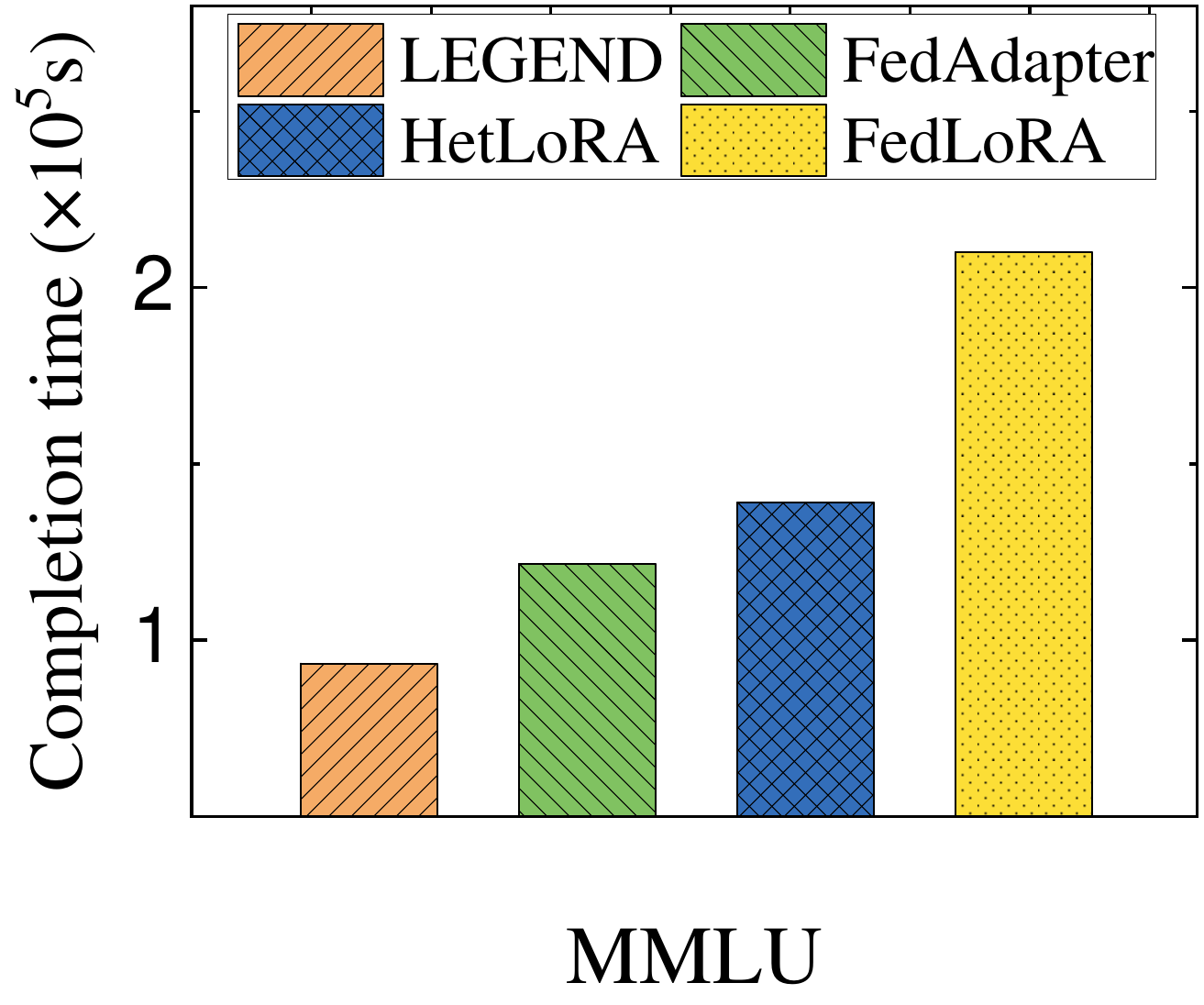}
        \end{minipage}%
        \label{llama2-7b-mmlu-completion}
    }%
    \centering
    \vspace{-0.5cm}
    \caption{\bluenote{Results on massive multitask understanding tasks.}}
    \label{mmlu-results}
    \vspace{-0.6cm}
\end{figure}

\bluenote{
2) \textbf{\textit{Massive Multitask Understanding}} 
aims to test model generalization and knowledge transfer by fine-tuning across multiple related tasks. 
We utilize the Massive Multitask Language Understanding (MMLU) datasets \cite{hendrycks2020measuring} for the evaluation, which is a comprehensive multiple-choice benchmark spanning 57 tasks across diverse academic domains. 
By sampling 20,000 training and 2,000 test samples from the \textit{auxiliary\_train} subset, we create a dataset that challenges models to demonstrate broad knowledge and nuanced reasoning capabilities.
We fine-tune a 7B-parameter Llama2 model (\ie, Llama2-7B) \cite{touvron2023llama}, which is composed of 32 transformer layers, on this customized dataset.
}

\bluenote{
3) \textbf{\textit{Mathematical Reasoning}}
serves as a critical task for evaluating the models' logical reasoning and computational reasoning capabilities. 
We employ the Grade School Math (GSM-8K) dataset \cite{cobbe2021training}, which is a widely recognized benchmark designed to evaluate models' elementary mathematical reasoning and problem-solving abilities. 
It comprises 7,473 training and 1,319 test problems, each requiring multi-step reasoning and arithmetic calculations, reflecting the complexity of grade-school mathematics. 
This dataset challenges models to demonstrate logical consistency, numerical computation, and interpretive skills essential for advanced mathematical understanding.
We fine-tune Llama2-7B on this dataset.
}


\textbf{Baselines.}
To evaluate the effectiveness of LEGEND, we adopt two LoRA-based approaches (\ie, naive FedLoRA \cite{zhang2023fedpetuning} and advanced HetLoRA \cite{cho2023heterogeneous}) and the state-of-the-art FedFT approaches (\ie, FedAdapter \cite{cai2022fedadapter}) as baselines.

1) \textbf{FedLoRA} 
integrates LoRA \cite{hu2021lora} into FedFT, where all the devices fine-tune the same local model with the identical rank applied to all transformer layers.

2) \textbf{HetLoRA}  
is an advanced LoRA-based approach for FedFT, which assigns each device with a diverse but appropriate LoRA rank for fine-tuning all transformer layers of its local model so as to deal with system heterogeneity.

3) \textbf{FedAdapter} 
is the state-of-the-art FedFT approach, which introduces Adapters \cite{houlsby2019parameter} in FedFT and dynamically searches for the optimal Adapter configuration to improve fine-tuning efficiency.

\textbf{Metrics.}
The following metrics are adopted to evaluate the performance of LEGEND and the baselines.

1) \textbf{\textit{Test Accuracy}} reflects the accuracy of the models fine-tuned by different approaches on the test datasets, measured by the proportion of correctly predicted data.
Specifically, we record the test accuracy of the global model (the model after aggregation at the PS) in each round.

2) \textit{\textbf{Completion time}} represents the total wall-clock time required for fine-tuning a model to achieve a target accuracy. 
For fair comparisons, we set the target accuracy as the minimum accuracy achieved by the four methods. 
We record the time of each round, summing it up to obtain the completion time, and also record the average waiting time to reflect the fine-tuning efficiency of different approaches.

3) \textit{\textbf{Communication Traffic}} is recorded by summing up the traffic for transmitting the trainable parameters between the PS and devices during model fine-tuning, which is used to measure the communication efficiency of each approach.

\textbf{Experimental Parameters.}
\bluenote{
By default, all experiments are carried out on our prototype system and run 100 rounds.
Each device fine-tunes 1 epoch per round using AdamW \cite{loshchilov2017decoupled} optimizer locally.
The learning rate is set as 0.002 and decays according to a cosine scheduler.
The batch size is fixed at 4 and the maximum sequence length is set to 512 for all experiments.
}
\vspace{-0.2cm}


\subsection{Overall Performance}
\begin{figure}[t]
    \hspace{-6mm}
    \centering
    \subfigure[Fine-tuning process (GSM-8K)]{
        \begin{minipage}[b]{0.5\linewidth}
        \centering
        \includegraphics[width=1.57in]{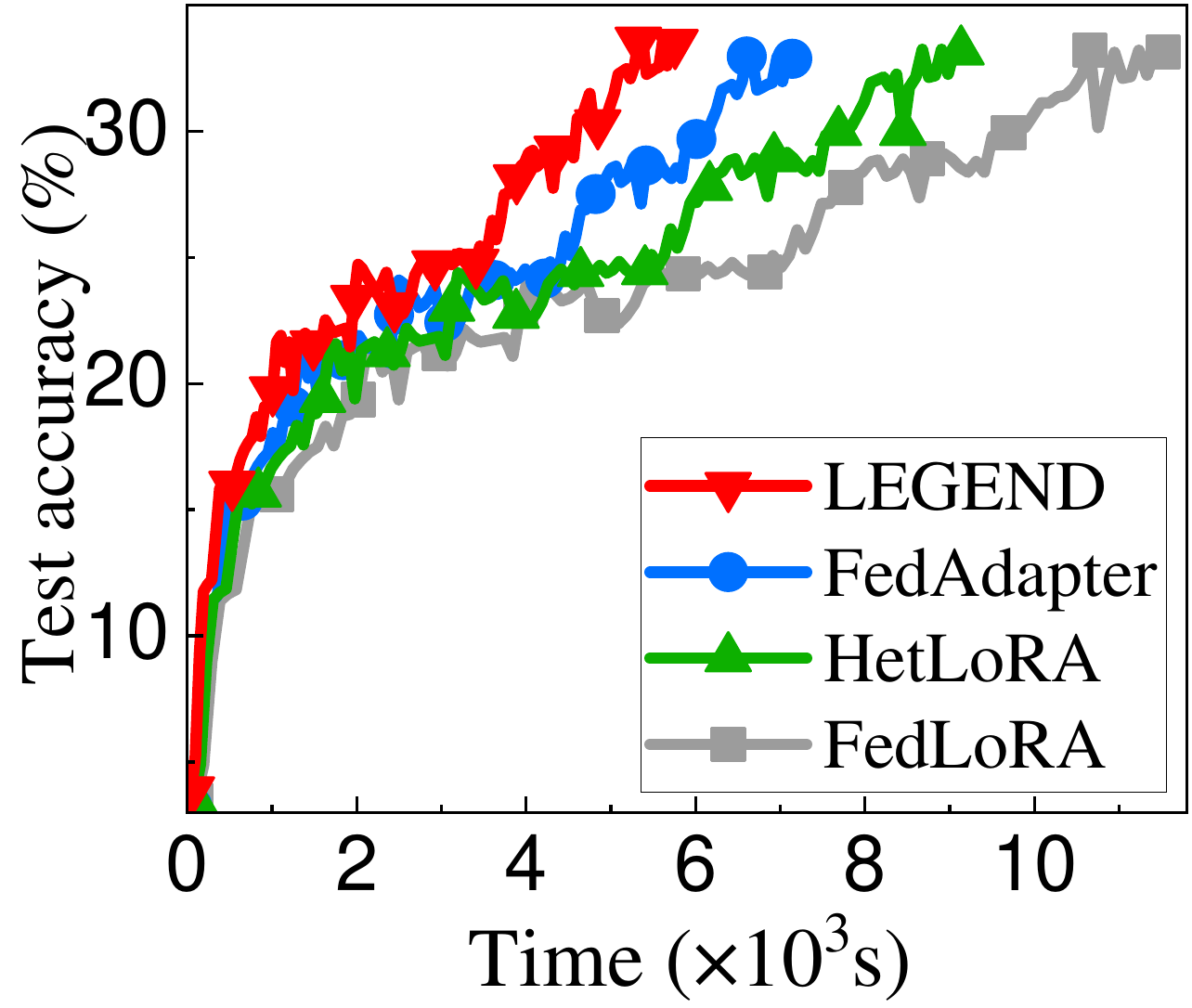}
        \end{minipage}%
        \label{llama2-7b-gsm8k-time2acc}
    }%
    \subfigure[Time to reach 30\% accuracy]{
        \begin{minipage}[b]{0.5\linewidth}
        \centering
        \includegraphics[width=1.6in]{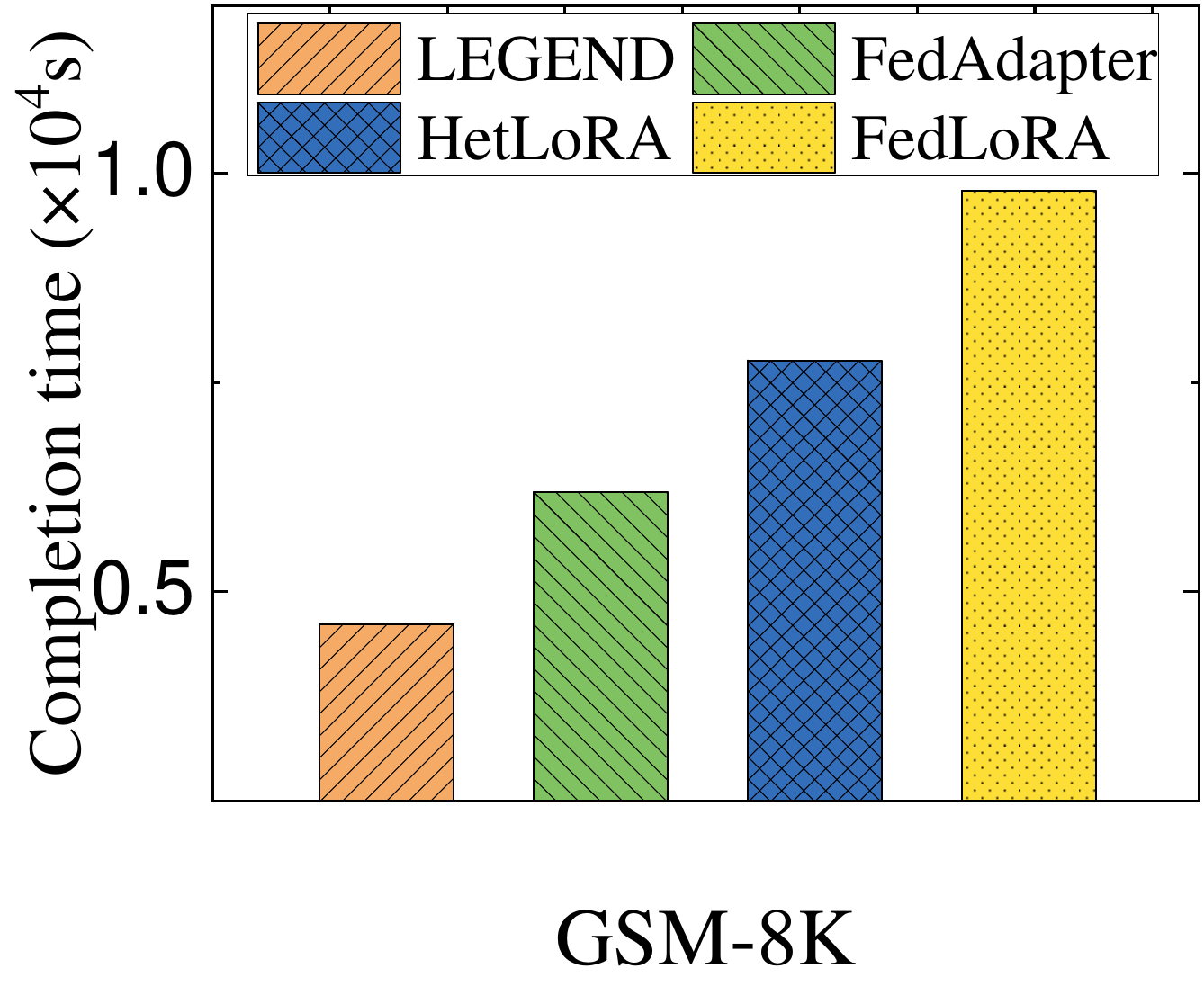}
        \end{minipage}%
        \label{llama2-7b-gsm8k-completion}
    }%
    \centering
    \vspace{-0.3cm}
    \caption{\bluenote{Results on mathematical reasoning tasks.}}
    \label{gsm8k-results}
    \vspace{-0.5cm}
\end{figure}

\vspace{-0.1cm}
Firstly, we conduct sets of experiments to evaluate the performance of LEGEND and the baselines.
The fine-tuning processes and the completion time of the general language understanding task are presented in Figures \ref{time-acc} and \ref{completion-time}, respectively.
By the results, LEGEND achieves the fastest convergence rate, outperforming the other approaches by a significant margin on all tasks. 
By assigning smaller LoRA depths with an optimized rank distribution for resource-constrained devices, LEGEND effectively enhances fine-tuning performance while reducing the time for local fine-tuning.
For instance, by Figures \ref{roberta-time2acc-sst2} and \ref{roberta-completion-sst2}, LEGEND takes only 1,479s to achieve 85\% accuracy on SST-2, while FedAdapter, HetLoRA, and FedLoRA take 2,412s, 2,503s, and 4,074s, respectively.
Compared to FedAdapter, HetLoRA, and FedLoRA, LEGEND provides 1.6$\times$, 1.7$\times$, and 2.8$\times$ speedup, respectively. 
By Figures \ref{roberta-time2acc-qnli} and \ref{roberta-completion-qnli}, LEGEND also outperforms the other baselines in terms of completion time for QNLI. 
Similarly, Figures \ref{deberta-time2acc-qqp} and \ref{deberta-completion-qqp} show that LEGEND takes 121,156s to achieve 87\% accuracy for QQP, while FedAdapter, HetLoRA, and FedLoRA consume 173,375s, 209,196s, and 269,749s, respectively. 
LEGEND separately achieves a speedup of 1.4$\times$, 1.7$\times$, and 2.2$\times$, compared with FedAdapter, HetLoRA, and FedLoRA.
For MNLI, as shown in Figures \ref{deberta-time2acc-mnli} and \ref{deberta-completion-mnli}, LEGEND speeds up the fine-tuning process by about 1.2$\times$, 1.5$\times$, and 2.1$\times$, compared to FedAdapter, HetLoRA, and FedLoRA, respectively. 
\bluenote{
Then, we present the experimental results for massive multitask understanding and mathematical reasoning tasks in Figures \ref{mmlu-results} and \ref{gsm8k-results}.
LEGEND consistently outperforms all other baselines, significantly accelerating the fine-tuning process. 
Specifically, as shown in Figure \ref{mmlu-results}, LEGEND achieves a speedup of 1.3$\times$, 1.5$\times$, and 2.3$\times$ compared to FedAdapter, HetLoRA, and FedLoRA, respectively. 
In Figure \ref{gsm8k-results}, LEGEND reduces the total completion time required to achieve 30\% accuracy by approximately 26\%, 41\%, and 54\% when compared to FedAdapter, HetLoRA, and FedLoRA, respectively.
These results demonstrate the superiority of LEGEND in accelerating the fine-tuning process through joint optimization of LoRA depth and rank distribution.
}

\begin{figure*}
	\centering
	\subfigure[Traffic to reach 95\% accuracy]{
		\begin{minipage}[b]{0.23\textwidth}
			\includegraphics[width=1\textwidth]{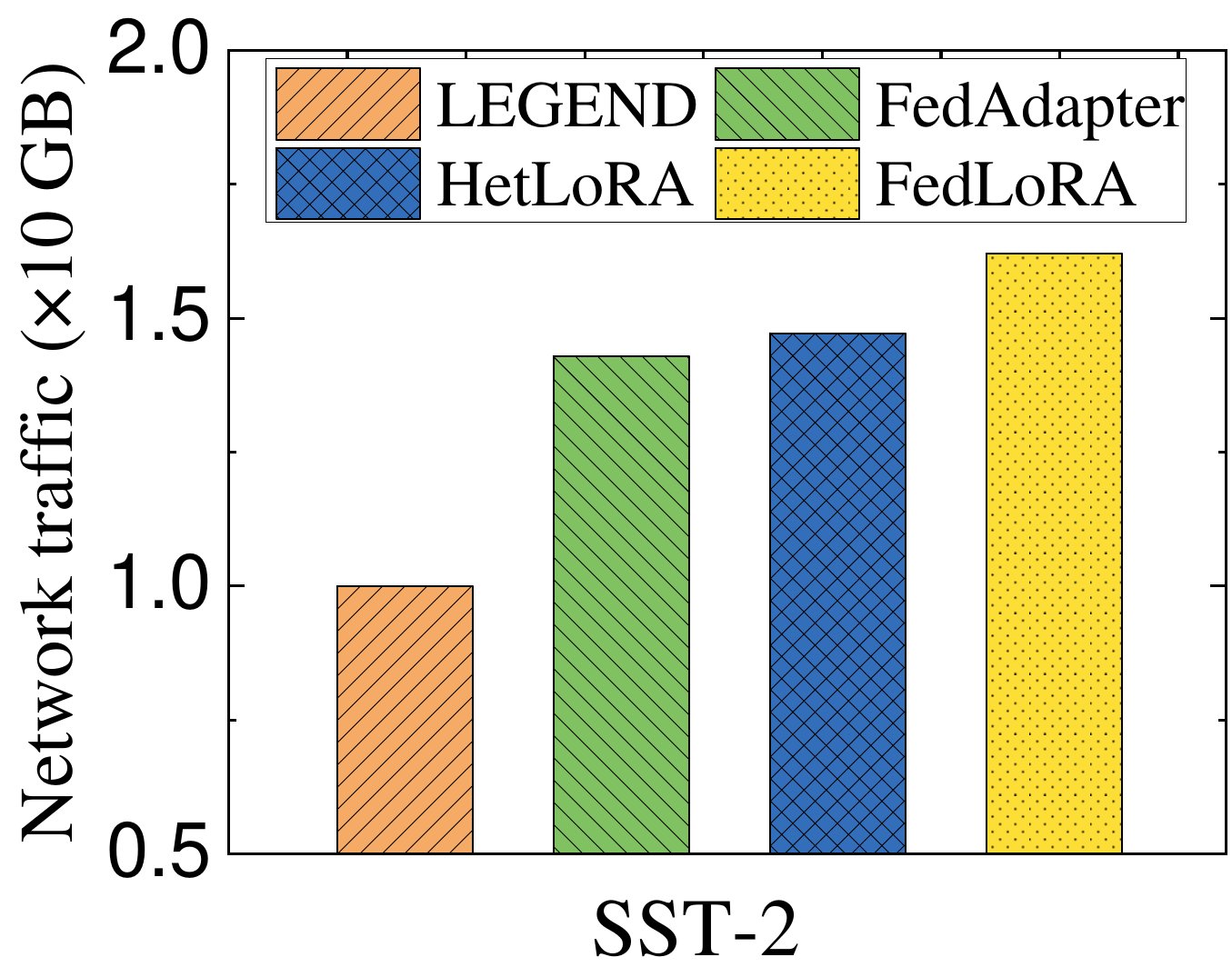} 
		\end{minipage}
		\label{roberta-traffic-sst2}
	}
    	\subfigure[Traffic to reach 90\% accuracy]{
    		\begin{minipage}[b]{0.23\textwidth}
   		 	\includegraphics[width=1\textwidth]{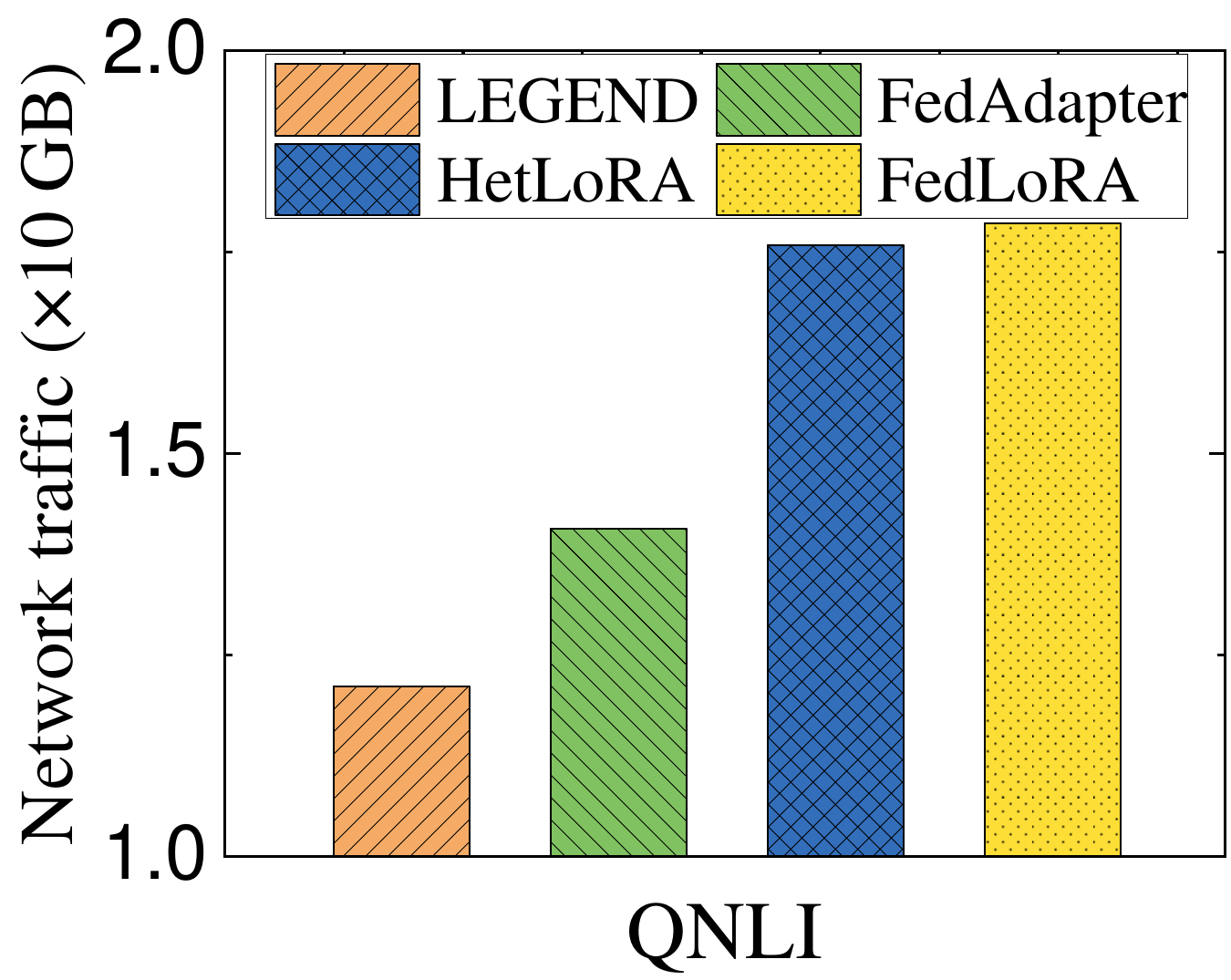}
    		\end{minipage}
		\label{roberta-traffic-qnli}
    	}
	\subfigure[Traffic to reach 87\% accuracy]{
		\begin{minipage}[b]{0.23\textwidth}
			\includegraphics[width=1\textwidth]{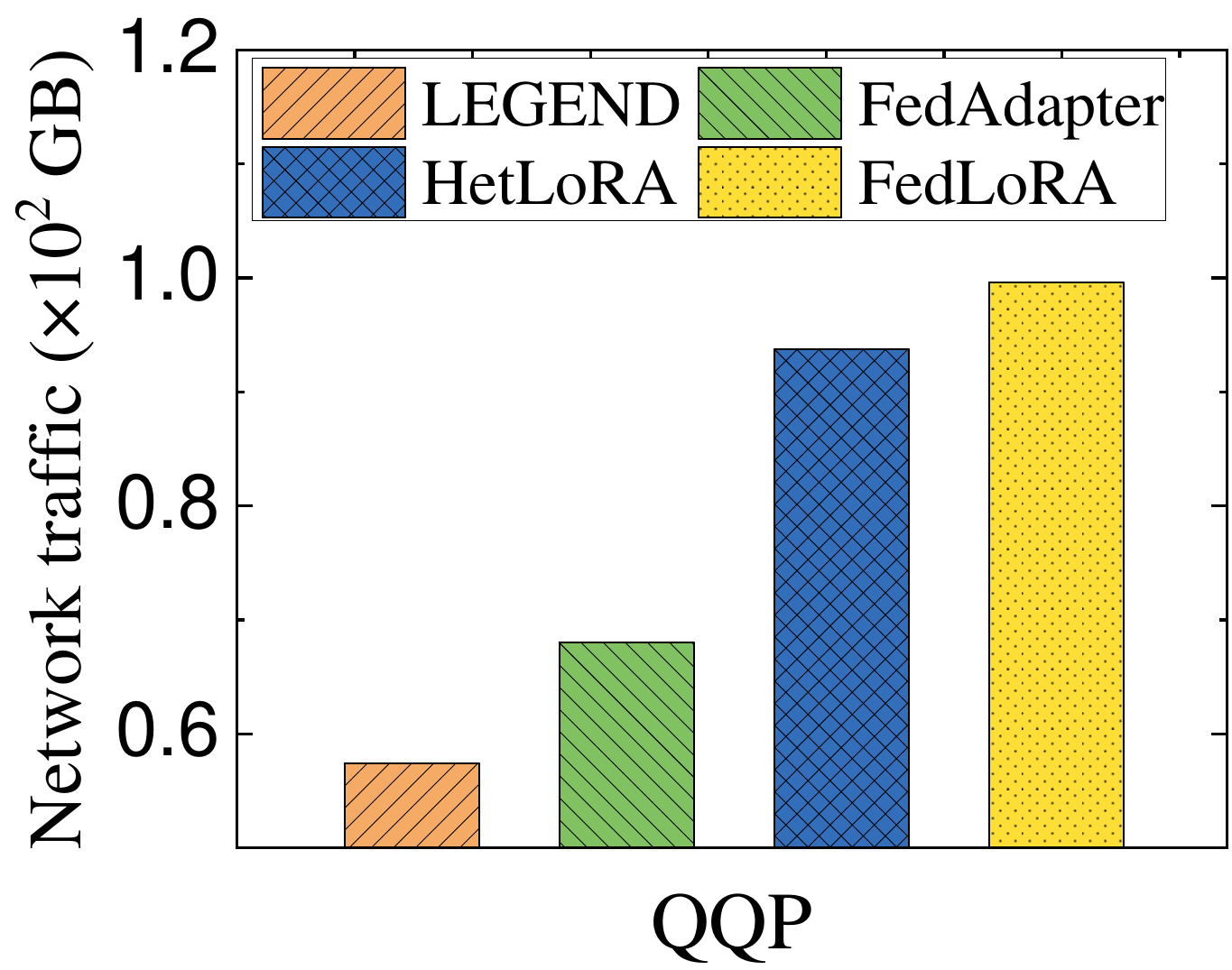} 
		\end{minipage}
		\label{deberta-traffic-qqp}
	}
    	\subfigure[Traffic to reach 85\% accuracy]{
    		\begin{minipage}[b]{0.23\textwidth}
		 	\includegraphics[width=1\textwidth]{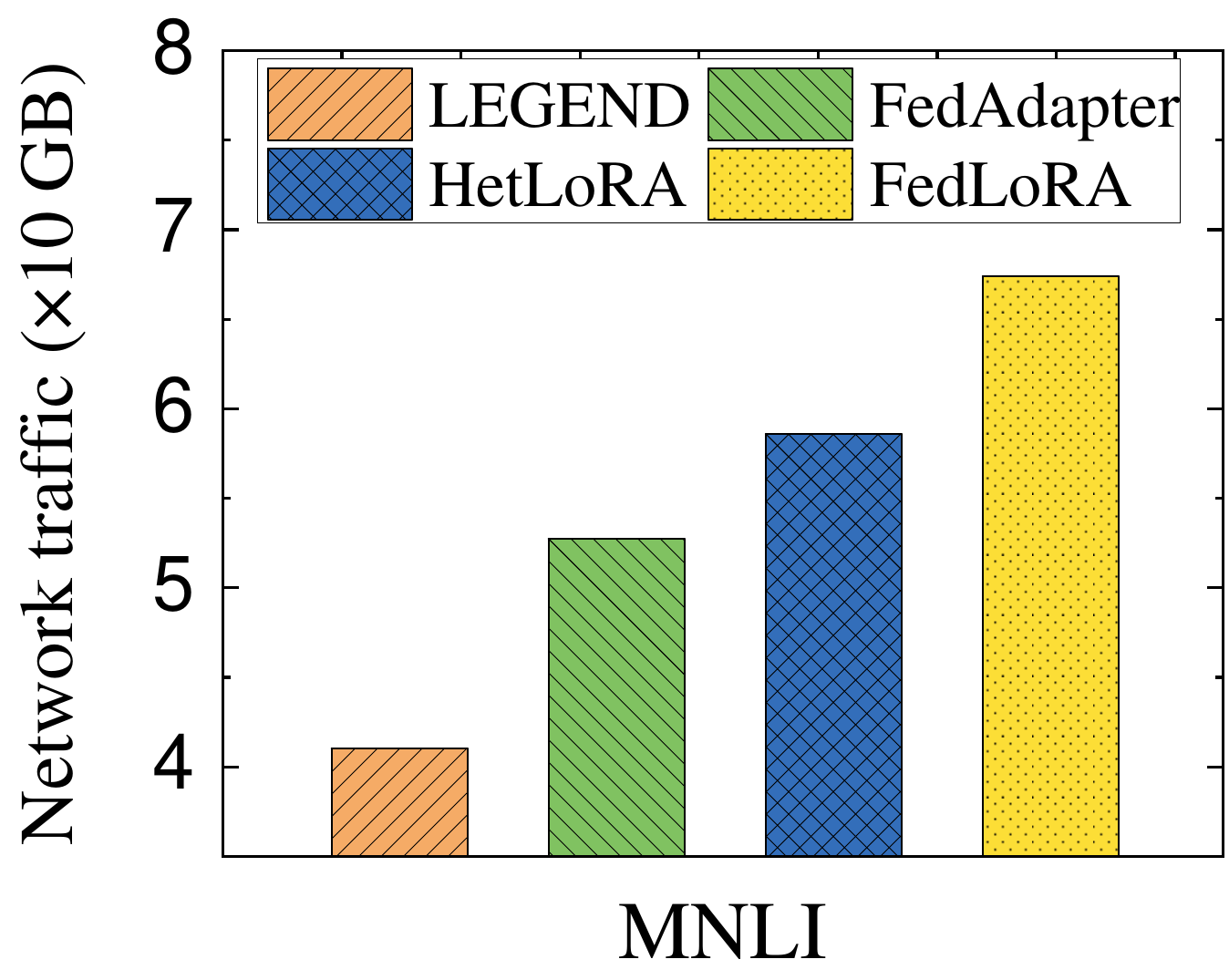}
    		\end{minipage}
		\label{deberta-traffic-mnli}
    	}
        \vspace{-0.3cm}
	\caption{Communication traffic of four approaches on general language understanding tasks.}
	\label{network-traffic}
        \vspace{-0.4cm}
\end{figure*}
\begin{figure*}
	\centering
	\subfigure[SST-2]{
		\begin{minipage}[b]{0.23\textwidth}
			\includegraphics[width=1\textwidth]{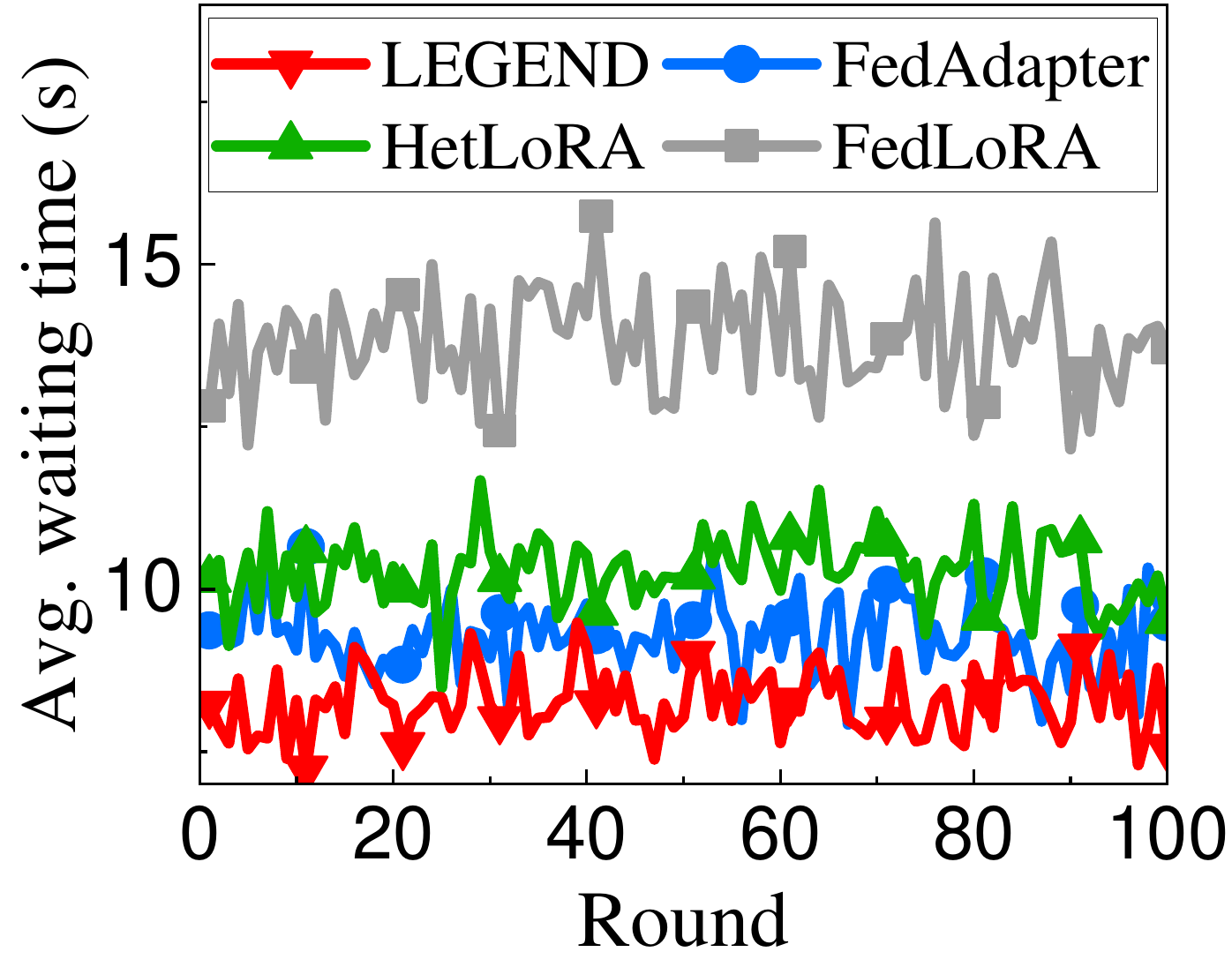} 
		\end{minipage}
		\label{roberta-waiting-sst2}
	}
    	\subfigure[QNLI]{
    		\begin{minipage}[b]{0.23\textwidth}
   		 	\includegraphics[width=1\textwidth]{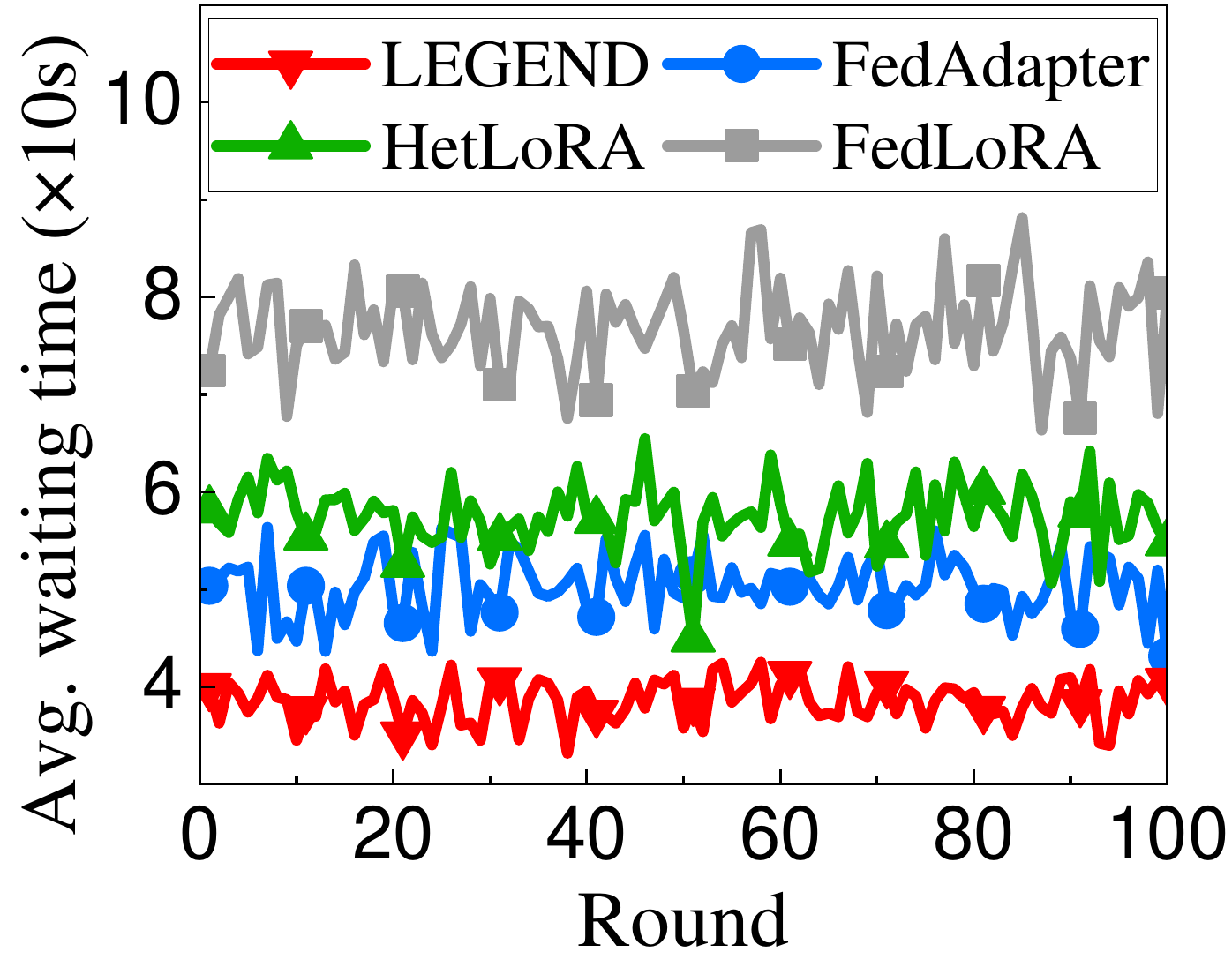}
    		\end{minipage}
		\label{roberta-waiting-qnli}
    	}
	\subfigure[QQP]{
		\begin{minipage}[b]{0.23\textwidth}
			\includegraphics[width=1\textwidth]{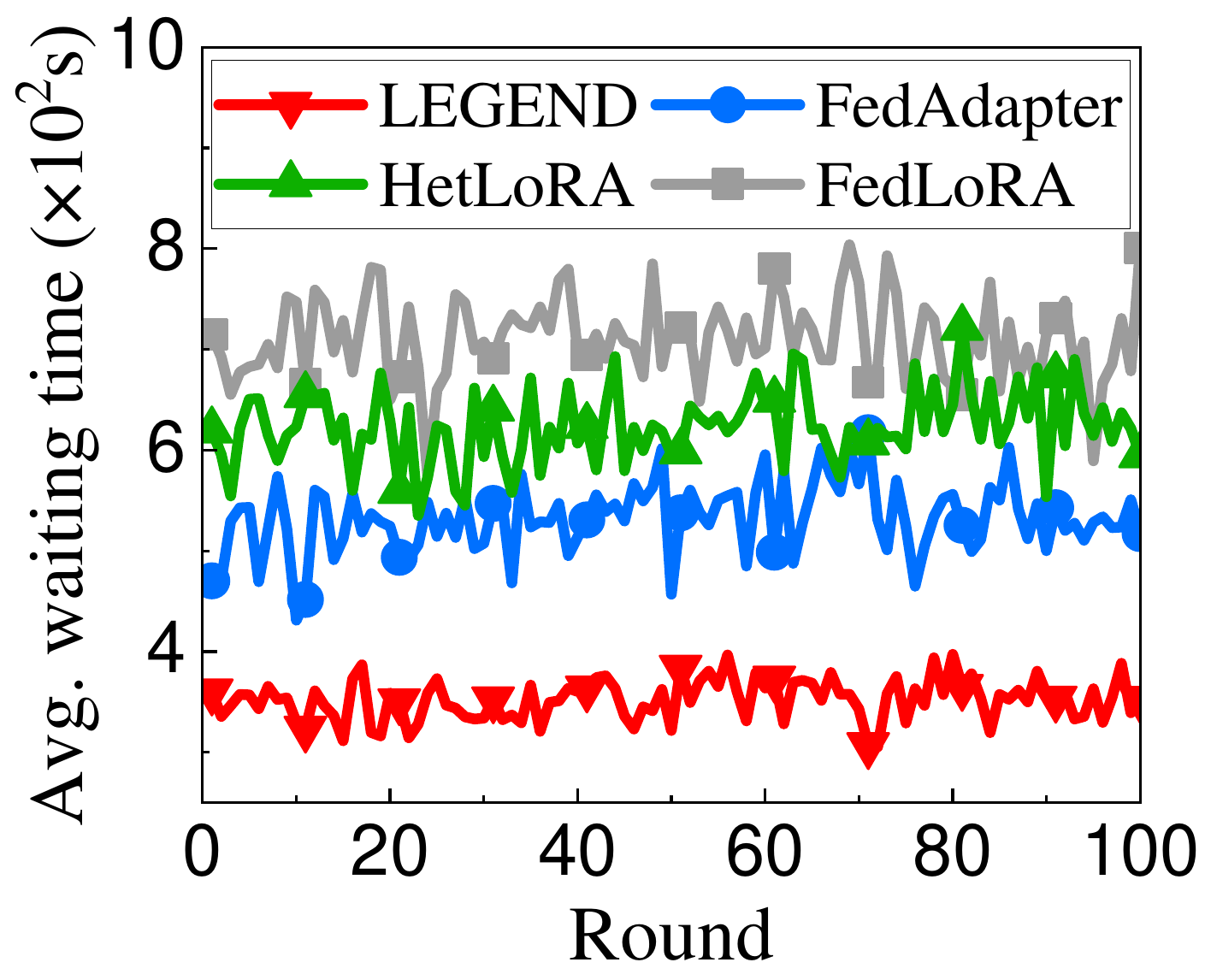} 
		\end{minipage}
		\label{deberta-waiting-qqp}
	}
    	\subfigure[MNLI]{
    		\begin{minipage}[b]{0.23\textwidth}
		 	\includegraphics[width=1\textwidth]{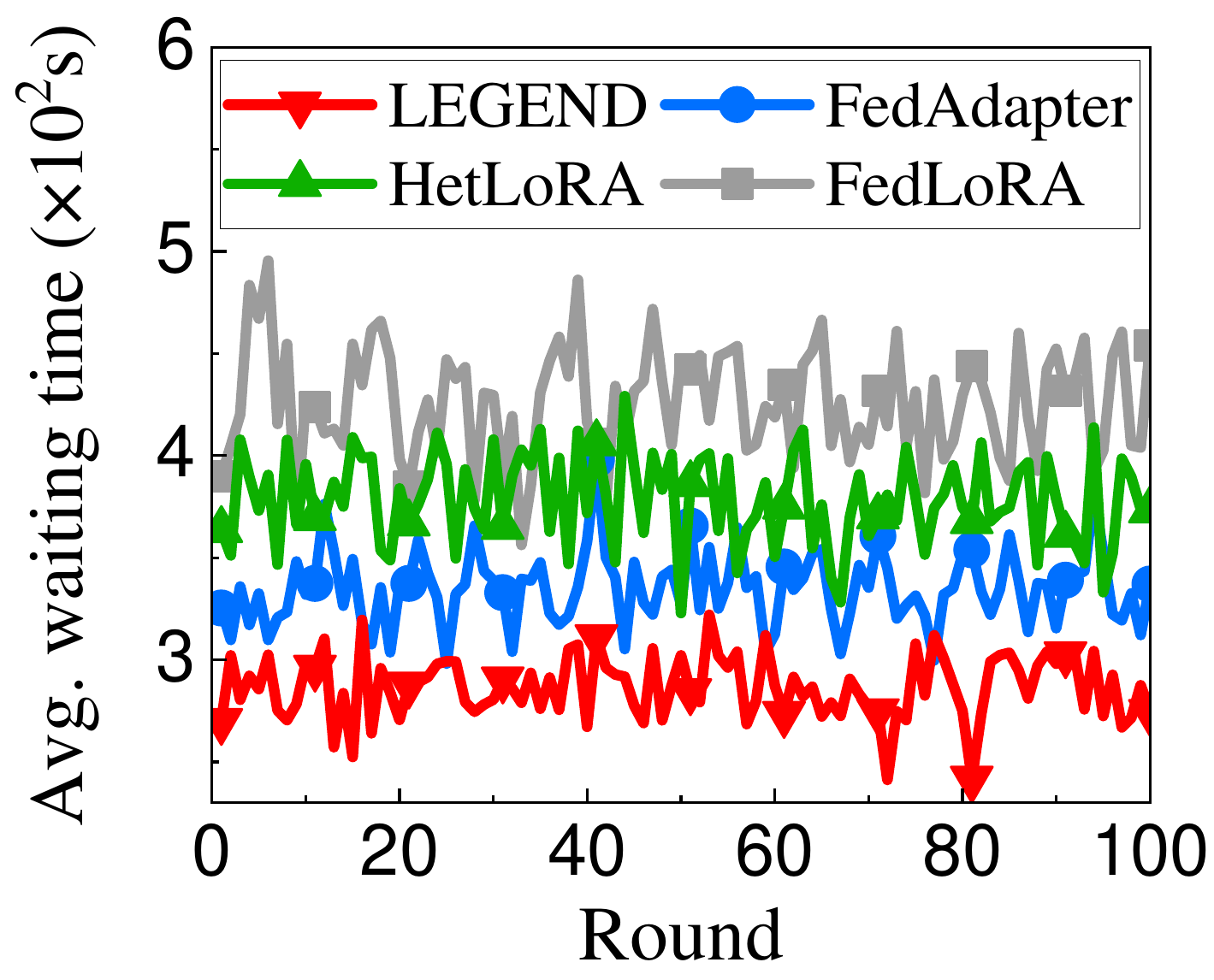}
    		\end{minipage}
		\label{deberta-waiting-mnli}
    	}
        \vspace{-0.3cm}
	\caption{Average waiting time of four approaches on general language understanding tasks.}
        \label{average-waiting-time}
        \vspace{-0.4cm}
\end{figure*}
Secondly, to illustrate the advantage of LEGEND in saving communication resources, we evaluate the communication traffic consumption of these approaches when achieving the target accuracy.
According to the results in Figure \ref{network-traffic}, LEGEND always causes the least communication traffic among all approaches. 
For example, as shown in Figure \ref{roberta-traffic-sst2}, when achieving 95\% accuracy, LEGEND consumes 9.9GB, while FedAdapter, HetLoRA, and FedLoRA consume 14.2GB, 14.7GB, and 16.2GB, respectively.
Compared with FedAdapter, HetLoRA, and FedLoRA, LEGEND reduces communication traffic by about 30.0\%, 32.1\%, and 38.3\%, respectively.
On one hand, compared to FedLoRA and HetLoRA which add LoRA layers of the same rank to all layers, LEGEND generates adaptive LoRA depth and rank distribution to accommodate heterogeneous devices. 
On the other hand, unlike FedAdapter which requires multiple device groups to find the optimal adapter structure, LEGEND determines the suitable LoRA configuration based on the device's status, making fine-tuning more efficient and thus reducing communication overhead.
Besides, by Figure \ref{roberta-traffic-qnli}, LEGEND consumes 12.1GB when achieving 90\% accuracy for QNLI, reducing the communication traffic by about 1.95GB, 5.46GB, and 5.74GB  compared with FedAdapter, HetLoRA, and FedLoRA, respectively.
For QQP, the results in Figure \ref{deberta-traffic-qqp} show that LEGEND reduces the total traffic consumption up to 42.3\% when achieving accuracy. 
Moreover, by Figure \ref{deberta-traffic-mnli}, when achieving 85\% accuracy for MNLI, LEGEND consumes 42GB, while FedAdapter, HetLoRA, and FedLoRA consume about 54GB, 60GB, and 69GB, respectively. 
\bluenote{
The experimental results underscore the advantages of LEGEND in reducing communication costs. 
}

To demonstrate the effectiveness of LEGEND towards system heterogeneity, we illustrate the average waiting time of the four approaches on the four datasets in Figure \ref{average-waiting-time}.
\bluenote{
LEGEND achieves the shortest waiting time across all datasets by assigning different LoRA depths with reasonable rank distribution to heterogeneous devices.
}
For example, by Figure \ref{roberta-waiting-sst2}, LEGEND separately improves the average waiting time by about 11.0\%, 19.5\%, and 40.2\%, compared with FedAdapter, HetLoRA, and FedLoRA for SST-2.
By Figure \ref{roberta-waiting-qnli}, the average waiting time of LEGEND is 38.6s for QNLI, while FedAdapter, HetLoRA, and FedLoRA incur average waiting time of 50.3s, 57.4s, and 76.7s, respectively.
LEGEND reduces the average waiting time by about 23.4\%, 32.8\%, and 49.7\% compared to FedAdapter, HetLoRA, and FedLoRA, respectively.
Besides, as illustrated by Figure \ref{deberta-waiting-qqp}, the average waiting time of LEGEND is 352s for SST-2 while FedAdapter, HetLoRA, and FedLoRA incur an average waiting time of 532s, 623s, and 710s, respectively.
Compared with FedAdapter, HetLoRA, and FedLoRA, LEGEND reduces the average waiting time by about 33.8\%, 43.5\%, and 50.4\%, respectively.
In addition, by Figure \ref{deberta-waiting-mnli}, LEGEND also outperforms FedAdapter, HetLoRA, and FedLoRA, reducing the averaging waiting time by 14\%, 24\%, and 32.7\%, respectively.
LEGEND achieves the least waiting time across all tasks by assigning appropriate depth to devices, significantly reducing the time for local fine-tuning on resource-constrained devices.
In contrast, FedLoRA adds LoRA layers to all transformer layers of the model, which requires the complete backpropagation process to update all LoRA layers and fast devices should be forced to wait for slow ones, leading to prolonged waiting time and poor fine-tuning efficiency.
HetLoRA improves FedLoRA by assigning diverse ranks to the devices to alleviate the challenge of system heterogeneity but with limited success, while FedAdapter speeds up the fine-tuning process by selecting the best fine-tuning group from groups with different Adapter configurations in each round.
These experimental results demonstrate the superiority of LEGEND in mitigating system heterogeneity.
\vspace{-0.3cm}


\vspace{-0.1cm}
\subsection{Ablation Study}
\vspace{-0.1cm}
\label{sec-ablation}
There are two key factors in LEGEND, \ie, LoRA depth and rank distribution, which are developed to enhance the performance of FedLoRA.
Herein, we conduct several sets of ablation experiments on SST-2 and QNLI to validate the effectiveness of the two essential factors. 
We adopt LEGEND, LEGEND without LoRA depth (denoted as LEGEND w/o LD), LEGEND without rank distribution (denoted as LEGEND w/o RD) as the baselines.
As illustrated in Figure \ref{ablation}, both LoRA depth and rank distribution are essential in LEGEND, but they contribute to the system in different ways.
\bluenote{
For instance, by Figures \ref{ablation-rank-time2acc-sst2} and \ref{ablation-rank-time2acc-qnli}, LEGEND w/o LD achieves similar final test accuracy, \eg, 95.3\% on SST-2 and 91.6\% on QNLI, compared with LEGEND, while LEGEND w/o RD slightly degrades to 94.6\% on SST-2 and 90.5\% on QNLI, respectively.
Besides, by Figure \ref{ablation-rank-time2acc-sst2}, when achieving a target accuracy of 94\% on SST-2, LEGEND saves the completion time by about 50.3\% and 54.5\% compared with LEGEND w/o RD and LEGEND w/o LD, respectively.
By Figure \ref{ablation-rank-time2acc-qnli}, compared with LEGEND w/o RD and LEGEND w/o LD, LEGEND separately speeds up fine-tuning by about 1.38$\times$ and 1.93$\times$, when achieving the same target accuracy (\ie, 90\% on QNLI). 
On the one hand, by employing a reasonable rank distribution and enabling all devices to fine-tune all LoRA layers, LEGEND w/o LD enhances the fine-tuning performance, but results in prolonged waiting time and slow convergence rates. 
On the other hand, LEGEND w/o RD assigns adaptive LoRA depth for the devices with uniform ranks across all layers, failing to strategically allocate larger rank to critical layers (\eg, deep layers), which constrains the fine-tuning performance and leads to performance degradation.
The results demonstrate the positive roles of the two designs.
}
\vspace{-0.2cm}

%% file: contents/related_word.tex
\begin{figure}[t]
    \centering
    \subfigure[SST-2]{
        \begin{minipage}[t]{0.5\linewidth}
        \centering
        \includegraphics[width=1.6in]{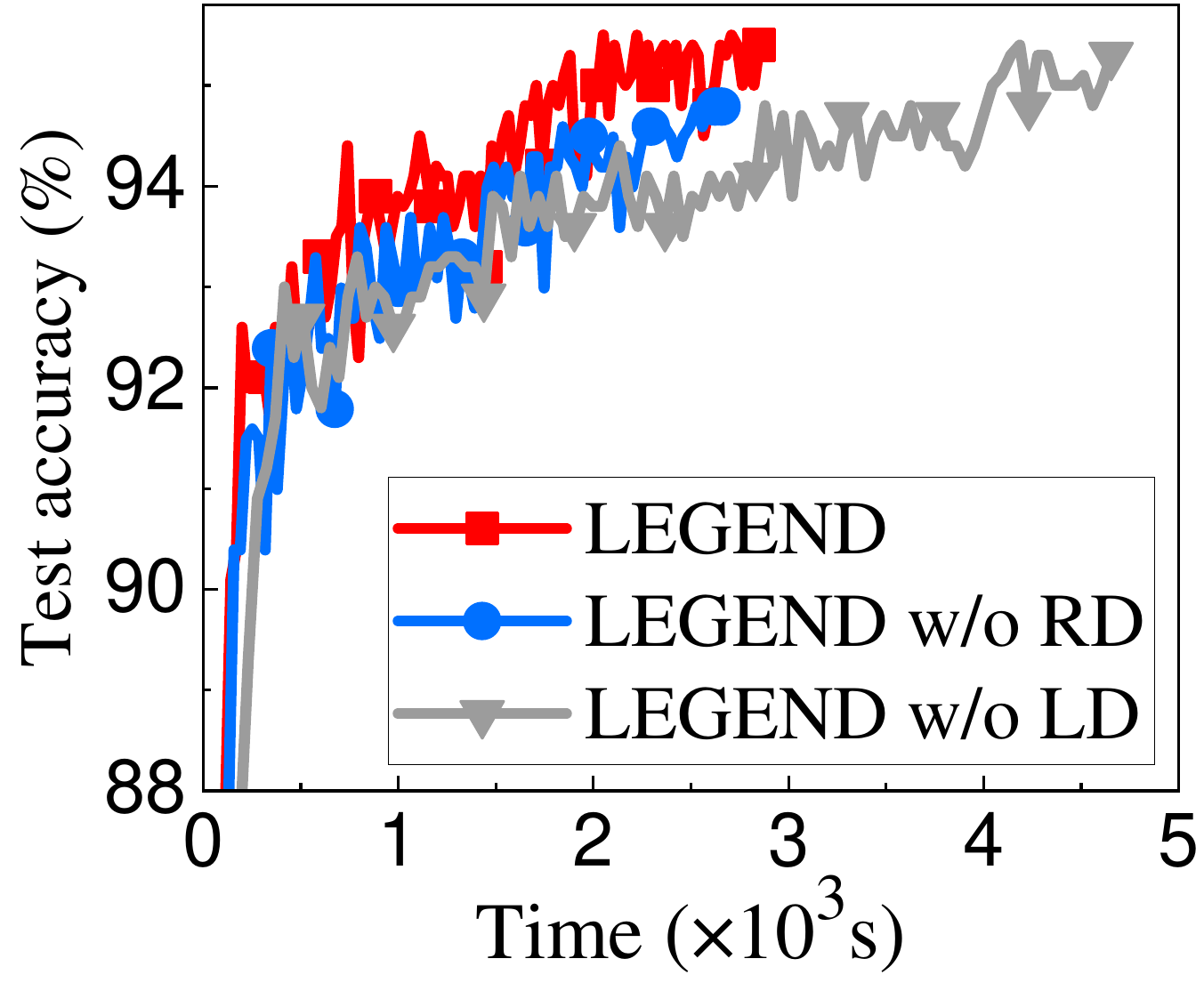}
        \end{minipage}%
        \label{ablation-rank-time2acc-sst2}
    }%
    \subfigure[QNLI]{
        \begin{minipage}[t]{0.5\linewidth}
        \centering
        \includegraphics[width=1.6in]{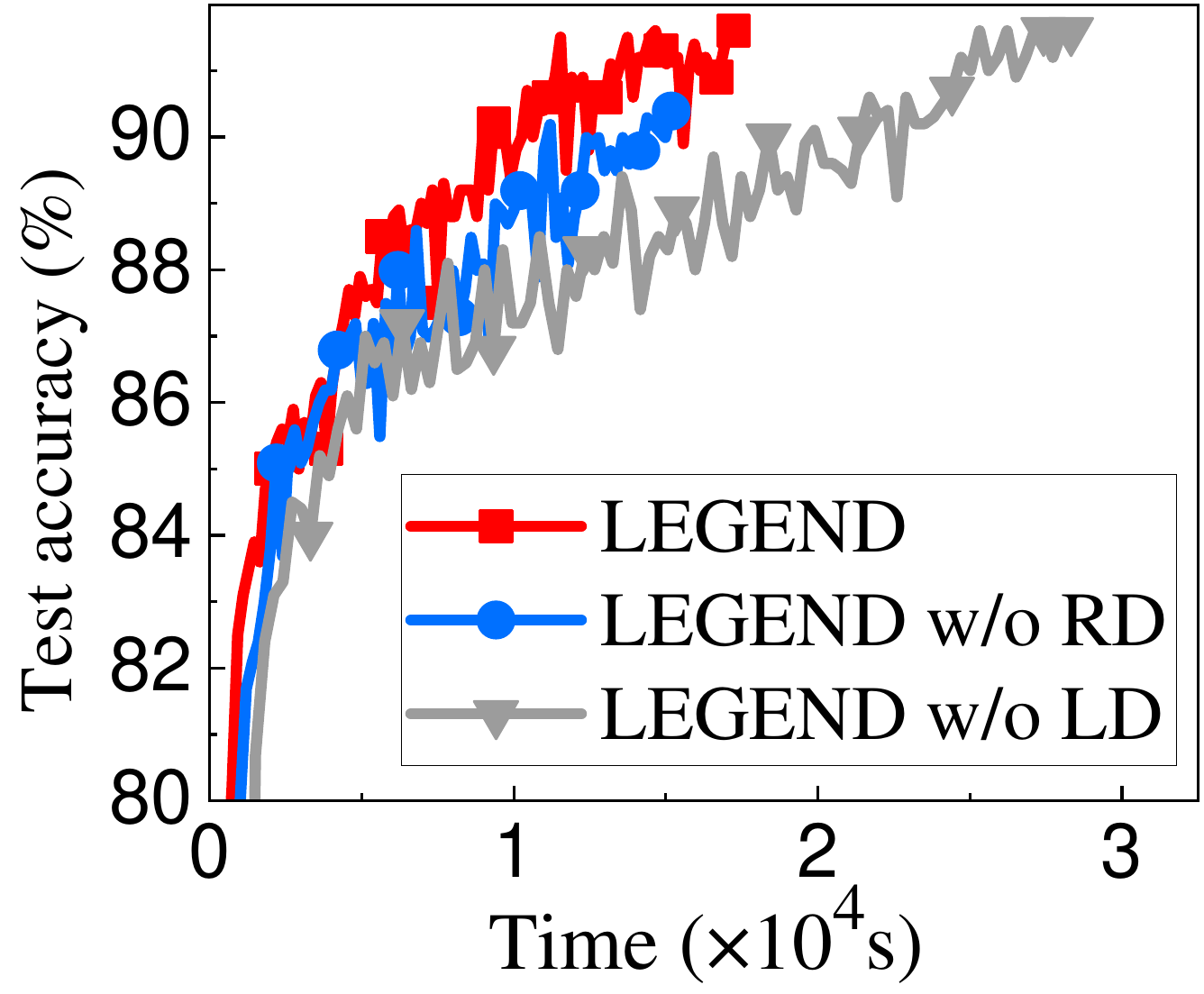}
        \end{minipage}%
        \label{ablation-rank-time2acc-qnli}
    }%
    \centering
    \vspace{-0.35cm}
    \caption{Effect of LoRA depth and rank distribution.}
    \label{ablation}
    \vspace{-0.4cm}
\end{figure}

\vspace{-0.2cm}
\textbf{Natural Language Processing.}
The development of modern natural language processing (NLP) traces back to the introduction of the transformer architecture by Vaswani \etal in 2017 \cite{vaswani2017attention}. 
Upon transformer architecture, groundbreaking language models, such as BERT \cite{devlin2018bert}, GPT-2 \cite{radford2019language}, and more recently Llama \cite{touvron2023llama}, have achieved state-of-the-art results across various NLP tasks. 
Specifically, the language model (LM) is first pre-trained on a large corpus to learn general features and patterns. 
Subsequently, the LM is further trained on domain-specific data generated on devices to enhance the model performance for a specific task \cite{devlin2018bert, brown2020language}. 
However, due to data privacy, it is impractical to collect enough data from the devices for centralized fine-tuning \cite{mcmahan2017communication, lin2021fednlp}. 

\textbf{Federated Fine-Tuning.} 
To fully utilize the massive data on devices, federated fine-tuning (FedFT) has been proposed to perform fine-tuning in a distributed manner \cite{lin2021fednlp}.
However, the high resource costs associated with FedFT pose significant challenges to its practical implementation.
Modern FedFT utilizes parameter-efficient fine-tuning (PEFT) methods, such as Adapter \cite{houlsby2019parameter, cai2022fedadapter}, prompt-tuning \cite{lester2021power, zhao2023fedprompt}, to reduce on-device resources costs. 
For example, Cai \etal \cite{cai2022fedadapter} first apply Adapter in FedFT and propose FedAdapter, which dynamically searches for the optimal Adapter structure to improve the fine-tuning efficiency.
However, due to the intrinsic property of Adapter, the fine-tuned Adapter-based LM inevitably brings additional inference latency, potentially resulting in up to a 30\% latency increase \cite{hu2021lora, liao2023parameter, li2024caraserve}, which is often unacceptable in practical applications. 
Upon prompt-tuning, Zhao \etal \cite{zhao2023fedprompt} propose FedPrompt to realize communication-efficient and privacy-preserving fine-tuning in federated setting.
Nevertheless, prompt-tuning unavoidably occupies a portion of the model's input length, consequently diminishing the usable input space and extra inference latency.

\textbf{Low-Rank Adaptation.}
To mitigate the disadvantages of the aforementioned PEFT methods, Hu \etal \cite{hu2021lora} propose low-rank adaptation (LoRA), which adds trainable rank decomposition matrices to each transformer layer of the LM while freezing the pre-trained weights of the LM to improve fine-tuning efficiency. 
LoRA achieves comparable performance of fully fine-tuning while introducing no additional inference latency, and has been widely adopted.
For instance, Dettmers \etal \cite{dettmers2024qlora} propose QLoRA, which is an efficient fine-tuning method based on model quantization, significantly reducing the memory requirement.
Chen \etal \cite{chen2023longlora} propose LongLoRA, which is an efficient fine-tuning approach to extend the context sizes of the LM with limited cost.


\textbf{FedFT with LoRA.}
LoRA is naturally incorporated into FedFT to reduce resource costs.
For example, Zhang \etal \cite{zhang2023fedpetuning} propose FedLoRA and verify the efficiency of LoRA in the context of FedFT through extensive experiments.
Upon FedLoRA, Cho \etal \cite{cho2023heterogeneous} propose HetLoRA, in which each device adds LoRA layers to all transformer layers with a diverse and appropriate LoRA rank to deal with system heterogeneity.
However, due to the rank mismatch of all LoRA layers on different devices, it is difficult to aggregate these layers, resulting in poor fine-tuning performance.
In conclusion, existing works simply add LoRA layers with the uniform rank distribution for all transformer layers, which still requires substantial computing and communication resources, resulting in slow fine-tuning speeds on weak devices. 
Moreover, system heterogeneity further leads to low fine-tuning efficiency or poor fine-tuning performance \cite{kim2021autofl, luo2022tackling}.
They do not address the challenges of resource constraints and system heterogeneity.
\vspace{-0.2cm}

%% file: contents/conclusion.tex
\vspace{-0.2cm}
In this paper, we review the intrinsic properties of FedFT and propose an efficient LoRA-based FedFT framework, called LEGEND, to address resource constraints and system heterogeneity.
We analyze the coupled relationship between LoRA depth and rank distribution, and design an efficient LoRA configuration algorithm for heterogeneous devices, thereby promoting fine-tuning efficiency.
Extensive experiments are conducted on a real platform of 80 wireless devices. 
The experimental results show that LEGEND significantly outperforms the existing methods, providing a speedup of 1.5-2.8$\times$ and saving communication costs by about 42.3\% while achieving the target accuracy. 

%% file: main.bbl
\begin{thebibliography}{10}

\bibitem{vaswani2017attention}
Ashish Vaswani, Noam Shazeer, Niki Parmar, Jakob Uszkoreit, Llion Jones,
  Aidan~N Gomez, {\L}ukasz Kaiser, and Illia Polosukhin.
\newblock Attention is all you need.
\newblock {\em Advances in neural information processing systems}, 30, 2017.

\bibitem{bai2020binarybert}
Haoli Bai, Wei Zhang, Lu~Hou, Lifeng Shang, Jing Jin, Xin Jiang, Qun Liu,
  Michael Lyu, and Irwin King.
\newblock Binarybert: Pushing the limit of bert quantization.
\newblock {\em arXiv preprint arXiv:2012.15701}, 2020.

\bibitem{devlin2018bert}
Jacob Devlin, Ming-Wei Chang, Kenton Lee, and Kristina Toutanova.
\newblock Bert: Pre-training of deep bidirectional transformers for language
  understanding.
\newblock {\em arXiv preprint arXiv:1810.04805}, 2018.

\bibitem{hou2020dynabert}
Lu~Hou, Zhiqi Huang, Lifeng Shang, Xin Jiang, Xiao Chen, and Qun Liu.
\newblock Dynabert: Dynamic bert with adaptive width and depth.
\newblock {\em Advances in Neural Information Processing Systems},
  33:9782--9793, 2020.

\bibitem{brown2020language}
Tom Brown, Benjamin Mann, Nick Ryder, Melanie Subbiah, Jared~D Kaplan, Prafulla
  Dhariwal, Arvind Neelakantan, Pranav Shyam, Girish Sastry, Amanda Askell,
  et~al.
\newblock Language models are few-shot learners.
\newblock {\em Advances in neural information processing systems},
  33:1877--1901, 2020.

\bibitem{mcmahan2017communication}
Brendan McMahan, Eider Moore, Daniel Ramage, Seth Hampson, and Blaise~Aguera
  y~Arcas.
\newblock Communication-efficient learning of deep networks from decentralized
  data.
\newblock In {\em Artificial intelligence and statistics}, pages 1273--1282.
  PMLR, 2017.

\bibitem{lin2021fednlp}
Bill~Yuchen Lin, Chaoyang He, Zihang Zeng, Hulin Wang, Yufen Huang, Christophe
  Dupuy, Rahul Gupta, Mahdi Soltanolkotabi, Xiang Ren, and Salman Avestimehr.
\newblock Fednlp: Benchmarking federated learning methods for natural language
  processing tasks.
\newblock {\em arXiv preprint arXiv:2104.08815}, 2021.

\bibitem{liao2023accelerating}
Yunming Liao, Yang Xu, Hongli Xu, Zhiwei Yao, Lun Wang, and Chunming Qiao.
\newblock Accelerating federated learning with data and model parallelism in
  edge computing.
\newblock {\em IEEE/ACM Transactions on Networking}, 2023.

\bibitem{stremmel2021pretraining}
Joel Stremmel and Arjun Singh.
\newblock Pretraining federated text models for next word prediction.
\newblock In {\em Advances in Information and Communication: Proceedings of the
  2021 Future of Information and Communication Conference (FICC), Volume 2},
  pages 477--488. Springer, 2021.

\bibitem{cai2022fedadapter}
Dongqi Cai, Yaozong Wu, Shangguang Wang, Felix~Xiaozhu Lin, and Mengwei Xu.
\newblock Fedadapter: Efficient federated learning for modern nlp.
\newblock {\em arXiv preprint arXiv:2205.10162}, 2022.

\bibitem{xu2022adaptive}
Yang Xu, Yunming Liao, Hongli Xu, Zhenguo Ma, Lun Wang, and Jianchun Liu.
\newblock Adaptive control of local updating and model compression for
  efficient federated learning.
\newblock {\em IEEE Transactions on Mobile Computing}, 22(10):5675--5689, 2022.

\bibitem{zhu2023pockengine}
Ligeng Zhu, Lanxiang Hu, Ji~Lin, Wei-Ming Chen, Wei-Chen Wang, Chuang Gan, and
  Song Han.
\newblock Pockengine: Sparse and efficient fine-tuning in a pocket.
\newblock In {\em Proceedings of the 56th Annual IEEE/ACM International
  Symposium on Microarchitecture}, pages 1381--1394, 2023.

\bibitem{dhar2021survey}
Sauptik Dhar, Junyao Guo, Jiayi Liu, Samarth Tripathi, Unmesh Kurup, and Mohak
  Shah.
\newblock A survey of on-device machine learning: An algorithms and learning
  theory perspective.
\newblock {\em ACM Transactions on Internet of Things}, 2(3):1--49, 2021.

\bibitem{liao2024mergesfl}
Yunming Liao, Yang Xu, Hongli Xu, Lun Wang, Zhiwei Yao, and Chunming Qiao.
\newblock Mergesfl: Split federated learning with feature merging and batch
  size regulation.
\newblock In {\em 2024 IEEE 40th International Conference on Data Engineering
  (ICDE)}, pages 2054--2067. IEEE, 2024.

\bibitem{touvron2023llama}
Hugo Touvron, Thibaut Lavril, Gautier Izacard, Xavier Martinet, Marie-Anne
  Lachaux, Timoth{\'e}e Lacroix, Baptiste Rozi{\`e}re, Naman Goyal, Eric
  Hambro, Faisal Azhar, et~al.
\newblock Llama: Open and efficient foundation language models.
\newblock {\em arXiv preprint arXiv:2302.13971}, 2023.

\bibitem{hu2021lora}
Edward~J Hu, Yelong Shen, Phillip Wallis, Zeyuan Allen-Zhu, Yuanzhi Li, Shean
  Wang, Lu~Wang, and Weizhu Chen.
\newblock Lora: Low-rank adaptation of large language models.
\newblock {\em arXiv preprint arXiv:2106.09685}, 2021.

\bibitem{liao2023adaptive}
Yunming Liao, Yang Xu, Hongli Xu, Lun Wang, and Chen Qian.
\newblock Adaptive configuration for heterogeneous participants in
  decentralized federated learning.
\newblock In {\em IEEE INFOCOM 2023-IEEE Conference on Computer
  Communications}, pages 1--10. IEEE, 2023.

\bibitem{lai2021oort}
Fan Lai, Xiangfeng Zhu, Harsha~V Madhyastha, and Mosharaf Chowdhury.
\newblock Oort: Efficient federated learning via guided participant selection.
\newblock In {\em 15th $\{$USENIX$\}$ Symposium on Operating Systems Design and
  Implementation ($\{$OSDI$\}$ 21)}, pages 19--35, 2021.

\bibitem{liu2023yoga}
Jun Liu, Jianchun Liu, Hongli Xu, Yunming Liao, Zhiyuan Wang, and Qianpiao Ma.
\newblock Yoga: Adaptive layer-wise model aggregation for decentralized
  federated learning.
\newblock {\em IEEE/ACM Transactions on Networking}, 2023.

\bibitem{zhang2023fedpetuning}
Zhuo Zhang, Yuanhang Yang, Yong Dai, Qifan Wang, Yue Yu, Lizhen Qu, and Zenglin
  Xu.
\newblock Fedpetuning: When federated learning meets the parameter-efficient
  tuning methods of pre-trained language models.
\newblock In {\em Annual Meeting of the Association of Computational
  Linguistics 2023}, pages 9963--9977. Association for Computational
  Linguistics (ACL), 2023.

\bibitem{houlsby2019parameter}
Neil Houlsby, Andrei Giurgiu, Stanislaw Jastrzebski, Bruna Morrone, Quentin
  De~Laroussilhe, Andrea Gesmundo, Mona Attariyan, and Sylvain Gelly.
\newblock Parameter-efficient transfer learning for nlp.
\newblock In {\em International conference on machine learning}, pages
  2790--2799. PMLR, 2019.

\bibitem{liao2023parameter}
Baohao Liao, Yan Meng, and Christof Monz.
\newblock Parameter-efficient fine-tuning without introducing new latency.
\newblock {\em arXiv preprint arXiv:2305.16742}, 2023.

\bibitem{li2024caraserve}
Suyi Li, Hanfeng Lu, Tianyuan Wu, Minchen Yu, Qizhen Weng, Xusheng Chen, Yizhou
  Shan, Binhang Yuan, and Wei Wang.
\newblock Caraserve: Cpu-assisted and rank-aware lora serving for generative
  llm inference.
\newblock {\em arXiv preprint arXiv:2401.11240}, 2024.

\bibitem{wankhade2022survey}
Mayur Wankhade, Annavarapu Chandra~Sekhara Rao, and Chaitanya Kulkarni.
\newblock A survey on sentiment analysis methods, applications, and challenges.
\newblock {\em Artificial Intelligence Review}, 55(7):5731--5780, 2022.

\bibitem{minaee2021deep}
Shervin Minaee, Nal Kalchbrenner, Erik Cambria, Narjes Nikzad, Meysam
  Chenaghlu, and Jianfeng Gao.
\newblock Deep learning--based text classification: a comprehensive review.
\newblock {\em ACM computing surveys (CSUR)}, 54(3):1--40, 2021.

\bibitem{zhang2022adaptive}
Qingru Zhang, Minshuo Chen, Alexander Bukharin, Pengcheng He, Yu~Cheng, Weizhu
  Chen, and Tuo Zhao.
\newblock Adaptive budget allocation for parameter-efficient fine-tuning.
\newblock In {\em The Eleventh International Conference on Learning
  Representations}, 2022.

\bibitem{cho2023heterogeneous}
Yae~Jee Cho, Luyang Liu, Zheng Xu, Aldi Fahrezi, Matt Barnes, and Gauri Joshi.
\newblock Heterogeneous lora for federated fine-tuning of on-device foundation
  models.
\newblock In {\em International Workshop on Federated Learning in the Age of
  Foundation Models in Conjunction with NeurIPS 2023}, 2023.

\bibitem{kim2021autofl}
Young~Geun Kim and Carole-Jean Wu.
\newblock Autofl: Enabling heterogeneity-aware energy efficient federated
  learning.
\newblock In {\em MICRO-54: 54th Annual IEEE/ACM International Symposium on
  Microarchitecture}, pages 183--198, 2021.

\bibitem{luo2022tackling}
Bing Luo, Wenli Xiao, Shiqiang Wang, Jianwei Huang, and Leandros Tassiulas.
\newblock Tackling system and statistical heterogeneity for federated learning
  with adaptive client sampling.
\newblock In {\em IEEE INFOCOM 2022-IEEE conference on computer
  communications}, pages 1739--1748. IEEE, 2022.

\bibitem{wikidump}
Wikimedia Foundation.
\newblock Wikimedia downloads.

\bibitem{2019t5}
Colin Raffel, Noam Shazeer, Adam Roberts, Katherine Lee, Sharan Narang, Michael
  Matena, Yanqi Zhou, Wei Li, and Peter~J. Liu.
\newblock Exploring the limits of transfer learning with a unified text-to-text
  transformer.
\newblock {\em arXiv e-prints}, 2019.

\bibitem{xu2023federated}
Mengwei Xu, Yaozong Wu, Dongqi Cai, Xiang Li, and Shangguang Wang.
\newblock Federated fine-tuning of billion-sized language models across mobile
  devices.
\newblock {\em arXiv preprint arXiv:2308.13894}, 2023.

\bibitem{loshchilov2017decoupled}
I~Loshchilov.
\newblock Decoupled weight decay regularization.
\newblock {\em arXiv preprint arXiv:1711.05101}, 2017.

\bibitem{kaddour2023challenges}
Jean Kaddour, Joshua Harris, Maximilian Mozes, Herbie Bradley, Roberta
  Raileanu, and Robert McHardy.
\newblock Challenges and applications of large language models.
\newblock {\em arXiv preprint arXiv:2307.10169}, 2023.

\bibitem{wei2024flexora}
Chenxing Wei, Yao Shu, Ying~Tiffany He, and Fei~Richard Yu.
\newblock Flexora: Flexible low rank adaptation for large language models.
\newblock {\em arXiv preprint arXiv:2408.10774}, 2024.

\bibitem{guo2019spottune}
Yunhui Guo, Honghui Shi, Abhishek Kumar, Kristen Grauman, Tajana Rosing, and
  Rogerio Feris.
\newblock Spottune: transfer learning through adaptive fine-tuning.
\newblock In {\em Proceedings of the IEEE/CVF conference on computer vision and
  pattern recognition}, pages 4805--4814, 2019.

\bibitem{zhang2023adaptive}
Qingru Zhang, Minshuo Chen, Alexander Bukharin, Pengcheng He, Yu~Cheng, Weizhu
  Chen, and Tuo Zhao.
\newblock Adaptive budget allocation for parameter-efficient fine-tuning.
\newblock {\em arXiv preprint arXiv:2303.10512}, 2023.

\bibitem{liu2019roberta}
Yinhan Liu, Myle Ott, Naman Goyal, Jingfei Du, Mandar Joshi, Danqi Chen, Omer
  Levy, Mike Lewis, Luke Zettlemoyer, and Veselin Stoyanov.
\newblock Roberta: A robustly optimized bert pretraining approach.
\newblock {\em arXiv preprint arXiv:1907.11692}, 2019.

\bibitem{wang2018glue}
Alex Wang, Amanpreet Singh, Julian Michael, Felix Hill, Omer Levy, and Samuel~R
  Bowman.
\newblock Glue: A multi-task benchmark and analysis platform for natural
  language understanding.
\newblock {\em arXiv preprint arXiv:1804.07461}, 2018.

\bibitem{song2024increasing}
Haobo Song, Hao Zhao, Soumajit Majumder, and Tao Lin.
\newblock Increasing model capacity for free: A simple strategy for parameter
  efficient fine-tuning.
\newblock {\em arXiv preprint arXiv:2407.01320}, 2024.

\bibitem{shuttleworth2024lora}
Reece Shuttleworth, Jacob Andreas, Antonio Torralba, and Pratyusha Sharma.
\newblock Lora vs full fine-tuning: An illusion of equivalence.
\newblock {\em arXiv preprint arXiv:2410.21228}, 2024.

\bibitem{ma2022layer}
Xiaosong Ma, Jie Zhang, Song Guo, and Wenchao Xu.
\newblock Layer-wised model aggregation for personalized federated learning.
\newblock In {\em Proceedings of the IEEE/CVF conference on computer vision and
  pattern recognition}, pages 10092--10101, 2022.

\bibitem{gao2024higher}
Chongyang Gao, Kezhen Chen, Jinmeng Rao, Baochen Sun, Ruibo Liu, Daiyi Peng,
  Yawen Zhang, Xiaoyuan Guo, Jie Yang, and VS~Subrahmanian.
\newblock Higher layers need more lora experts.
\newblock {\em arXiv preprint arXiv:2402.08562}, 2024.

\bibitem{konecny2016federated}
Jakub Konecn{\`y}, H~Brendan McMahan, Felix~X Yu, Peter Richt{\'a}rik,
  Ananda~Theertha Suresh, and Dave Bacon.
\newblock Federated learning: Strategies for improving communication
  efficiency.
\newblock {\em arXiv preprint arXiv:1610.05492}, 8, 2016.

\bibitem{leroy2019federated}
David Leroy, Alice Coucke, Thibaut Lavril, Thibault Gisselbrecht, and Joseph
  Dureau.
\newblock Federated learning for keyword spotting.
\newblock In {\em ICASSP 2019-2019 IEEE international conference on acoustics,
  speech and signal processing (ICASSP)}, pages 6341--6345. IEEE, 2019.

\bibitem{halperin2010predictable}
Daniel Halperin, Wenjun Hu, Anmol Sheth, and David Wetherall.
\newblock Predictable 802.11 packet delivery from wireless channel
  measurements.
\newblock {\em ACM SIGCOMM computer communication review}, 40(4):159--170,
  2010.

\bibitem{yue2017linkforecast}
Chaoqun Yue, Ruofan Jin, Kyoungwon Suh, Yanyuan Qin, Bing Wang, and Wei Wei.
\newblock Linkforecast: Cellular link bandwidth prediction in lte networks.
\newblock {\em IEEE Transactions on Mobile Computing}, 17(7):1582--1594, 2017.

\bibitem{zhang2022federated}
Zhuo Zhang, Yuanhang Yang, Yong Dai, Lizhen Qu, and Zenglin Xu.
\newblock When federated learning meets pre-trained language models'
  parameter-efficient tuning methods.
\newblock {\em arXiv preprint arXiv:2212.10025}, 2022.

\bibitem{mittal2019survey}
Sparsh Mittal.
\newblock A survey on optimized implementation of deep learning models on the
  nvidia jetson platform.
\newblock {\em Journal of Systems Architecture}, 97:428--442, 2019.

\bibitem{merkel2014docker}
Dirk Merkel et~al.
\newblock Docker: lightweight linux containers for consistent development and
  deployment.
\newblock {\em Linux j}, 239(2):2, 2014.

\bibitem{naik2016building}
Nitin Naik.
\newblock Building a virtual system of systems using docker swarm in multiple
  clouds.
\newblock In {\em 2016 IEEE International Symposium on Systems Engineering
  (ISSE)}, pages 1--3. IEEE, 2016.

\bibitem{paszke2019pytorch}
Adam Paszke, Sam Gross, Francisco Massa, Adam Lerer, James Bradbury, Gregory
  Chanan, Trevor Killeen, Zeming Lin, Natalia Gimelshein, Luca Antiga, et~al.
\newblock Pytorch: An imperative style, high-performance deep learning library.
\newblock {\em Advances in neural information processing systems}, 32, 2019.

\bibitem{gabriel2004open}
Edgar Gabriel, Graham~E Fagg, George Bosilca, Thara Angskun, Jack~J Dongarra,
  Jeffrey~M Squyres, Vishal Sahay, Prabhanjan Kambadur, Brian Barrett, Andrew
  Lumsdaine, et~al.
\newblock Open mpi: Goals, concept, and design of a next generation mpi
  implementation.
\newblock In {\em Recent Advances in Parallel Virtual Machine and Message
  Passing Interface: 11th European PVM/MPI Users’ Group Meeting Budapest,
  Hungary, September 19-22, 2004. Proceedings 11}, pages 97--104. Springer,
  2004.

\bibitem{tirumala1999iperf}
Iperf: The tcp/udp bandwidth measurement tool.
\newblock {\em http://dast. nlanr. net/Projects/Iperf/}, 1999.

\bibitem{wolf2020transformers}
Thomas Wolf, Lysandre Debut, Victor Sanh, Julien Chaumond, Clement Delangue,
  Anthony Moi, Pierric Cistac, Tim Rault, R{\'e}mi Louf, Morgan Funtowicz,
  et~al.
\newblock Transformers: State-of-the-art natural language processing.
\newblock In {\em Proceedings of the 2020 conference on empirical methods in
  natural language processing: system demonstrations}, pages 38--45, 2020.

\bibitem{he2021debertav3}
Pengcheng He, Jianfeng Gao, and Weizhu Chen.
\newblock Debertav3: Improving deberta using electra-style pre-training with
  gradient-disentangled embedding sharing.
\newblock {\em arXiv preprint arXiv:2111.09543}, 2021.

\bibitem{hendrycks2020measuring}
Dan Hendrycks, Collin Burns, Steven Basart, Andy Zou, Mantas Mazeika, Dawn
  Song, and Jacob Steinhardt.
\newblock Measuring massive multitask language understanding.
\newblock {\em arXiv preprint arXiv:2009.03300}, 2020.

\bibitem{wang2017deep}
Yan Wang, Xiaojiang Liu, and Shuming Shi.
\newblock Deep neural solver for math word problems.
\newblock In {\em Proceedings of the 2017 conference on empirical methods in
  natural language processing}, pages 845--854, 2017.

\bibitem{cobbe2021training}
Karl Cobbe, Vineet Kosaraju, Mohammad Bavarian, Mark Chen, Heewoo Jun, Lukasz
  Kaiser, Matthias Plappert, Jerry Tworek, Jacob Hilton, Reiichiro Nakano,
  et~al.
\newblock Training verifiers to solve math word problems.
\newblock {\em arXiv preprint arXiv:2110.14168}, 2021.

\bibitem{radford2019language}
Alec Radford, Jeffrey Wu, Rewon Child, David Luan, Dario Amodei, Ilya
  Sutskever, et~al.
\newblock Language models are unsupervised multitask learners.
\newblock {\em OpenAI blog}, 1(8):9, 2019.

\bibitem{lester2021power}
Brian Lester, Rami Al-Rfou, and Noah Constant.
\newblock The power of scale for parameter-efficient prompt tuning.
\newblock {\em arXiv preprint arXiv:2104.08691}, 2021.

\bibitem{zhao2023fedprompt}
Haodong Zhao, Wei Du, Fangqi Li, Peixuan Li, and Gongshen Liu.
\newblock Fedprompt: Communication-efficient and privacy-preserving prompt
  tuning in federated learning.
\newblock In {\em ICASSP 2023-2023 IEEE International Conference on Acoustics,
  Speech and Signal Processing (ICASSP)}, pages 1--5. IEEE, 2023.

\bibitem{dettmers2024qlora}
Tim Dettmers, Artidoro Pagnoni, Ari Holtzman, and Luke Zettlemoyer.
\newblock Qlora: Efficient finetuning of quantized llms.
\newblock {\em Advances in Neural Information Processing Systems}, 36, 2024.

\bibitem{chen2023longlora}
Yukang Chen, Shengju Qian, Haotian Tang, Xin Lai, Zhijian Liu, Song Han, and
  Jiaya Jia.
\newblock Longlora: Efficient fine-tuning of long-context large language
  models.
\newblock {\em arXiv preprint arXiv:2309.12307}, 2023.

\end{thebibliography}
